\documentclass[11pt,preprint]{aastex}
\newcommand{\cm}{\,{\rm cm}}
\newcommand{\GHz}{\,{\rm GHz}}
\newcommand{\gtsim}{\gtrsim}                
\newcommand{\kJy}{\,{\rm kJy}}
\newcommand{\K}{\,{\rm K}}
\newcommand{\kms}{\,{\rm km\,s^{-1}}}
\newcommand{\ltsim}{\lesssim}               
\newcommand{\s}{\,{\rm s}}
\newcommand{\sr}{\,{\rm sr}}
\newcommand{\yr}{\,{\rm yr}}
\newcommand{\odd}{{\rm o}}
\newcommand{\pc}{{\,\rm pc}}
\newcommand{\beq}{\begin{equation}}
\newcommand{\eeq}{\end{equation}}
\newcommand{\beqa}{\begin{eqnarray}}
\newcommand{\eeqa}{\end{eqnarray}}

\newcommand{\fhot}{f_{\rm H\alpha}^{(\rm hot)}}
\newcommand{\fcooling}{f_{\rm H\alpha}^{(\rm cooling)}}
\newcommand{\fHI}{f_{\rm H\alpha}^{(\rm H\,I)}}
\newcommand{\frefl}{f_{\rm H\alpha}^{(\rm refl)}}

\newcommand{\hb}{H$\beta$}
\newcommand{\hi}{\ion{H}{1}}
\newcommand{\hii}{\ion{H}{2}}
\newcommand{\nii}{[\ion{N}{2}]}
\newcommand{\sii}{[\ion{S}{2}]}

\newcommand{\oi}{[\ion{O}{1}]}
\newcommand{\oii}{[\ion{O}{2}]}

\newcommand{\oiii}{[\ion{O}{3}]}
\newcommand{\neii}{[\ion{Ne}{2}]}
\newcommand{\neiii}{[\ion{Ne}{3}]}
\newcommand{\hei}{\ion{He}{1}}

\newcommand{\nH}{{n_{\rm H}}}

\newcommand{\jphysb}{J Phys B}
\newcommand{\nature}{Nature}

\begin{document}
\title{H$\alpha$ and Free-Free Emission from the WIM}

\shorttitle{H$\alpha$ and Free-Free Emission from the WIM}

\shortauthors{Dong \& Draine}

\author{Ruobing Dong and B. T. Draine}

\affil{Department of Astrophysical Sciences, Princeton University,
Princeton, NJ, 08544}

\begin{abstract}
Recent observations have found the ratio of H$\alpha$ to free-free
radio continuum to be surprisingly high in the diffuse ionized ISM
(the so-called WIM), corresponding to an electron temperature of only
$\sim$3000~K.  Such low temperatures were unexpected in gas that was
presumed to be photoionized.  We consider a 3-component model for the
observed diffuse emission, consisting of a mix of (1) photoionized
gas, (2) gas that is recombining and cooling, and (3) cool H~I gas.
This model can successfully reproduce the observed intensities of
free-free continuum, H$\alpha$, and collisionally-excited lines such
as [\ion{N}{2}]6583.  To reproduce the low observed value of free-free
to H$\alpha$, the PAH abundance in the photoionized regions must be
lowered by a factor $\sim$3, and $\sim$20\% of the diffuse H$\alpha$
must be reflected from dust grains, as suggested by
\citet{Wood+Reynolds_1999}. 
\end{abstract}

\keywords{atomic processes, ISM}

\section{INTRODUCTION}\label{sec:introduction}

Low density ionized regions, often referred to as
the warm ionized medium (WIM), account for $\sim$90\% or more of the ionized
interstellar hydrogen in the Galaxy. 
The WIM occupies a substantial volume fraction
and is a major component of the interstellar medium (ISM)
at $\sim$1~kpc above the galactic disk plane in both the Milky Way
\citep{
McKee_1990,
Reynolds_1991,
Reynolds_1993,
Ferriere_2001}, and
in external galaxies
\citep{
Rand+Kulkarni+Hester_1990,
Ferguson+Wyse+Gallagher+Hunter_1996,
Hoopes+Walterbos+Greenwalt_1996,
Rand_1996,
Hoopes+Walterbos+Rand_1999,
Rossa+Dettmar_2000,
Collins+Rand_2001}.
Although it is now generally believed that only massive O-type stars
have enough ionizing power to account for the ionization in the WIM
\citep{Reynolds_2004}, the detailed ionization mechanism is still
not fully understood. Structures such as superbubbles and chimneys
have been invoked to explain how H-ionizing photons could
travel from the stellar sources to ionize the WIM.

Observations of optical diagnostic lines such as H$\alpha$, \hb, \hei$\lambda5876$,
\nii$\lambda6548, 6583, 5755$, \sii$\lambda6716, 6731$, \oi$\lambda6300$,
\oii$\lambda3727$, and \oiii$\lambda5007$, as
well as infrared lines (\neii12.81$~\mu$m and \neiii15.55$~\mu$m) have been 
made in the last two decades, aiming to understand the
physical conditions in the WIM 
\citep{
Greenawalt+Walterbos+Braun_1997, 
Rand_1997,
Rand_2000,
Haffner+Reynolds+Tufte_1999,
Otte+Gallagher+Reynolds_2002,
Miller+Veilleux_2003b,
Madsen+Reynolds+Haffner_2006}.
Although the observed line ratios vary from region to region,
optical line ratios such as \nii$\lambda6548$/H$\alpha$ appear to indicate that
the WIM material has temperature $T\approx 8000\pm2000\K$.
\citet{Hill+Benjamin+Kowal+etal_2008} estimate the characteristic
electron density in the WIM to be $n_e\approx0.07$cm$^{-3}$, 
with a volume filling fraction of $\sim$30\%. 
Optical line ratios such as \oiii$\lambda5007$/H$\alpha$ 
indicate that compared with \hii~regions powered by early O-type stars, 
the WIM seems to have fewer ions present that require ionization energies 
greater than
23~eV, with $n(\rm He^+)$/$n_{\rm He}$ ranging from 0.3 to 0.6
\citep{
Reynolds_1985,
Rand_1997,
Haffner+Reynolds+Tufte_1999,
Reynolds_2004,
Madsen+Reynolds+Haffner_2006,
Haffner+Dettmar+Beckman+etal_2009}.

Recently, \citet[hereafter DDF09]{Dobler+Draine+Finkbeiner_2009},
using the {\em WMAP} five-year data, 
deduced a surprisingly low temperature ($\sim$3000\,K) for the WIM
from the measured ratio of the thermal bremsstrahlung (free-free)
emission to the associated H$\alpha$ emission.
Similar results had previously been reported based on the free-free 
to H$\alpha$ ratio in the {\em WMAP}
one-year data \citep{Davies+Dickinson+Banday+etal_2006} and
three-year data \citep{Dobler+Finkbeiner_2008b}. 
The low value came as a surprise, 
because the diffuse ionized gas had been expected to have $T\approx8000$K,
based on both observed optical line ratios and the predictions of steady-state
photoionization models.

Previous discussions of the temperature and ionization of the WIM have
generally assumed steady-state conditions.
Here we instead argue that much of the WIM is {\it not} in
equilibrium, and that it is possible to understand the observed line ratios
and weak free-free emission if part of the WIM is assumed to be gas in 
the process of cooling and recombining
after removal of a photoionizing source,
as a consequence of either stellar evolution or changes in the opacity
along the sightline between the parcel of gas and the O star providing
Lyman continuum photons.
We propose that observations of free-free and line emission
from the diffuse ISM at high latitudes typically sum over
multiple components: 
(1) gas currently being photoionized, 
(2) gas that was photoionized in the past but which is currently 
cooling and recombining,
and (3) neutral gas (the cold neutral medium and warm neutral medium).
Dust grains mixed with the gas also contribute scattered
light that includes emission lines from H\,II regions in the disk.
For plausible choices of
parameters, we find that this model can account for the low ratio of
free-free to H$\alpha$ found by DDF09, while also reproducing
observed optical line ratios, such as [N\,II]6583/H$\alpha$.

The structure of this paper is as follows. In
Section~\ref{sec:model} we describe the model used to simulate
cooling and recombination in the WIM.
The simulation results are presented in Section~\ref{sec:results}, where we
explore various factors which affect this process. In
section~\ref{sec:threegasmodel} we propose a model using three ISM components
to explain the
observed line ratios and free-free emission. We discuss the main
result and our models in Section~\ref{sec:discussion}, followed by a
brief summary in Section~\ref{sec:summary}.

\section{Model Description}\label{sec:model}

We model the evolution of temperature, ionization, and emission from
cooling, recombining gas. As initial conditions we take the ionization
and temperature to be consistent with photoionization by a distant OB
association, but at $t=0$ we assume that the radiation with $h\nu >
13.6$~eV is suddenly turned off. The remaining
ionization and heating sources are cosmic rays and photoelectric emission by the
diffuse interstellar radiation field with $h\nu < 13.6$~eV.
We follow the ionization of 11 elements: H, He, C, N, O, Ne, Mg, Si, S,
Ar, and Fe, which includes all the elements that normally have gas phase
abundance relative to H above $10^{-6}$, and includes all important
coolants.


\subsection{Physics in the Model}\label{sec:phy-model}

The evolution is assumed to take place at constant volume, with
H nucleon density $\nH=const$.
The abundance $x_{A,r}\equiv n(A^{+r})/\nH$ 
of ion $A^{+r}$ evolves according to
\beqa\nonumber
\frac{dx_{A,r}}{dt} &=& \zeta_{A,r-1}x_{A,r-1} + 
\left[ n_e\left(\alpha_{A,r+1}^{(rr)}+\alpha_{A,r+1}^{(dr)}\right)
 +\nH\alpha_{A,r+1}^{(gr)}\right]x_{A,r+1}
\\
&&-\left[\zeta_{A,r}+n_e\left(\alpha_{A,r}^{(rr)}+\alpha_{A,r}^{(dr)}\right)
+\nH\alpha_{A,r}^{(gr)}\right]x_{A,r} \hspace*{1.0cm}{\rm for~}r\geq 1~,
\\
\frac{dx_{A,0}}{dt}
&=& \left[n_e\left(\alpha_{A,1}^{(rr)}+\alpha_{A,1}^{(de)}\right)+\nH\alpha_{A,1}^{(gr)}\right]x_{A,1}
-\zeta_{A,0}x_{A,0}~~~,
\eeqa
where $\zeta_{A,r}$ is the probability per unit time of ionization
$A^{+r}\rightarrow A^{+r+1}+e^-$
due to either photoionization, cosmic rays, or secondary electrons;
$\alpha_{A,r}^{(rr)}$ is the rate coefficient for radiative recombination;
$\alpha_{A,r}^{(dr)}$ is the rate coefficient for dielectronic recombination;
and $\alpha_{A,r}^{(gr)}$ is the effective rate coefficient
for recombination on dust grains.
The present model includes only ions with $r\leq2$, and
gas temperatures $T\leq 10^4\K$.
We assume case B
recombination for H and He. The recombination rate for H is 
taken to be \citep{Draine_2011}
\beq
\alpha_B=2.54\times10^{-13}T_4^{-0.8163-0.0208\ln{T_4}}\cm^3\s^{-1}
~~~,
\eeq
where $T_4\equiv T/10^4$K, and
for recombination of \ion{He}{2}$\rightarrow$\ion{He}{1} we take
\beq
\alpha_B({\rm He})=2.72\times10^{-13}T_4^{-0.789}\cm^3\s^{-1} ~~~.
\eeq

For other elements, 
radiative recombination rates are evaluated by subroutine 
rrfit.f from \citet{Verner_1999},
which uses rates from 
\citet{Pequignot+Petitjean+Boisson_1991} for
ions of C, N, O, and Ne
(refitted with the formula of \citet{Verner+Ferland_1996}),
and 
from \citet{Shull+vanSteenberg_1982} for ions of Mg, Si, S, Ar, and Fe.
Rate coefficients $\alpha^{(dr)}$
for Mg and Si were
from
\citet{Nussbaumer+Storey_1986},
for S were from \citet{Shull+vanSteenberg_1982},
and for Fe were from \citet{Arnaud+Raymond_1992}.
Rate coefficients $\alpha_{A,r}^{(gr)}$
for recombination on grains are taken from
\citet{Weingartner+Draine_2001d}.
The electron density
$n_e\equiv \nH\sum_A\sum_r rx_{A,r}$,
and the free particle density $n=n_e+\nH\sum_A\sum_r x_{A,r}$.

For isochoric evolution, the temperature $T$ evolves according to
\beq
\frac{dT}{dt} = \frac{\Gamma-\Lambda}{(3/2)nk} - 
T \frac{1}{n}\frac{dn}{dt} ~~~,
\eeq
where $(\Gamma-\Lambda)$ is the net rate of change of thermal kinetic
energy per unit volume due to heating
and cooling processes (including ionization and recombination), and
$dn/dt$ is the net rate of change of the free particle density
due to
ionization and recombination processes.
The heating rate per volume $\Gamma$ includes heat deposition by cosmic
ray ionization and by photoelectrons emitted from atoms, ions, and grains.
The cooling rate per volume $\Lambda$ includes kinetic energy removed
from the gas by inelastic collisions and by recombination of ions and electrons.
The mean kinetic energy per recombining electron is taken to be
\citep[][eq.\ 27.23]{Draine_2011}
\beq
\langle E_{\rm rr}\rangle \approx \left[0.684-0.0416\ln T_4\right]kT ~~~.
\eeq 
Radiative cooling processes
consist of free-free emission, and line emission following
collisional excitation of various atoms and ions.
Collisional deexcitation is included, 
although it is unimportant at the densities considered here.

Grain-related processes are included in our simulation.
From their study of the ``spinning dust" emission from the WIM,
DDF09 concluded that the PAH abundance in the WIM is a factor $\sim$3
lower than in the \hi. Because the PAHs account for a large fraction
of the photoelectric heating
\citep{Bakes+Tielens_1994,Weingartner+Draine_2001c}, 
and are also thought to dominate the
grain-assisted recombination
\citep{Weingartner+Draine_2001d}, depletion of the PAHs
will affect the heating and recombination.
To explore this, we multiply the rates for dust photoelectric heating
and grain-assisted recombination by a factor $g$, where $g=1$ gives
the rates estimated for normal grain abundances in the diffuse ISM
for photoelectric heating 
and grain-assisted recombination \citep{Weingartner+Draine_2001d}.
Our standard model for the WIM
assumes $g=1/3$, which presumably results primarily from
reduced abundances of the small grains that account for most of
the grain surface area. We will explore the sensitivity to the reduction factor $g$
by also performing simulations with $g=1$ (no reduction in PAH abundance)
and $g=0.1$, a factor of 10 suppression of grain photoelectric
heating and grain-assisted recombination.
Grain photoelectric heating is calculated following 
\citet{Weingartner+Draine_2001c}, and the $g=1$ rates for
grain-assisted recombination are taken from
\citet{Weingartner+Draine_2001d}.

We calculate collisional excitation and resulting radiative cooling
by ions of C, N, O, Ne, Si, S, Ar, and Fe.
We include a total of 159 lines, but the 26 
lines of listed in Table \ref{tab:cooling lines}
account for more than 95\% of the radiative cooling at each point in the
thermal evolution of our models.  Sources of the collisional
rate coefficients for species used in our calculations are listed in
Table \ref{tab:refs for rate coeffs}.
\begin{table}[h]
\begin{center}
\caption{\label{tab:cooling lines}
         Principal Cooling Lines (wavelengths {\it in vacuo})}
\begin{tabular}{l l l l}
\hline
{[\ion{C}{2}]}157.7$\micron$ &[\ion{O}{1}]145.5$\micron$  &[\ion{S}{2}]6733\AA\        &[\ion{Fe}{2}]25.99$\micron$\\
{[\ion{C}{2}]}2328\AA\       &[\ion{O}{1}]63.19$\micron$  &[\ion{S}{2}]6718\AA\        &[\ion{Fe}{2}]5.340$\micron$\\
{[\ion{C}{2}]}2326\AA\       &[\ion{O}{1}]6302\AA\        &[\ion{S}{2}]4070\AA\        &[\ion{Fe}{2}]1.644$\micron$\\
{[\ion{N}{2}]}205.3$\micron$ &[\ion{O}{2}]3730\AA\        &[\ion{S}{3}]33.48$\micron$ &[\ion{Fe}{2}]1.321$\micron$\\
{[\ion{N}{2}]}121.8$\micron$ &[\ion{O}{2}]3727\AA\        &[\ion{S}{3}]18.71$\micron$ &[\ion{Fe}{2}]1.257$\micron$\\
{[\ion{N}{2}]}6585\AA\       &[\ion{Ne}{2}]12.81$\micron$ &[\ion{S}{3}]9533\AA\ \\
{[\ion{N}{2}]}6550\AA\       &[\ion{Si}{2}]34.81$\micron$ &[\ion{S}{3}]9071\AA\ \\
\hline
\end{tabular}
\end{center}
\end{table}
\begin{table}[h]
\begin{center}
\caption{\label{tab:refs for rate coeffs}
         Sources for Collisional Rate Coefficients}
\begin{tabular}{l c l l}
\hline
Ion & Transition & $e^-$ & H$^0$\\
\hline
\ion{C}{2}& $^2{\rm P}_{1/2}^\odd- {^2{\rm P}}_{3/2}^\odd$ & \citet{Tayal_2008b}
                                                  & \citet{Barinovs+vanHemert+Krems+Dalgarno_2005}\\
\ion{N}{2}& $^3{\rm P}_J- {^3{\rm P}}_{J^\prime}$          & \citet{Hudson+Bell_2005} & ---\\
\ion{N}{2}& $^3{\rm P}_J- {^1{\rm D}}_2$                   & \citet{Hudson+Bell_2005} & ---\\
\ion{O}{1}& $^3{\rm P}_J- {^3{\rm P}}_{J^\prime}$           & \citet{Pequignot_1996} 
                                                  & \citet{Abrahamsson+Krems+Dalgarno_2007}\\
\ion{O}{1}& $^3{\rm P}_J- {^1{\rm D}}_2$                    & \citet{Pequignot_1996} & ---\\
\ion{O}{2}& $^4{\rm S}_{3/2}^\odd- {^2{\rm D}}_J^\odd$     & \citet{Tayal_2007} & ---\\
\ion{Ne}{2}& $^2{\rm P}_{3/2}^\odd- {^2{\rm P}}_{1/2}^\odd$& \citet{Griffin+Mitnik+Badnell_2001} & --- \\
\ion{Si}{2}& $^2{\rm P}_{1/2}^\odd- {^2{\rm P}}_{3/2}^\odd$& \citet{Bautista+Quinet+Palmeri+etal_2009} & ---\\
\ion{S}{2}& $^4{\rm S}_{3/2}^\odd- {^2{\rm D}}_J^\odd$     & \citet{Tayal+Zatsarinny_2010}
                                                  & \citet{Barinovs+vanHemert+Krems+Dalgarno_2005}\\
\ion{S}{3}& $^3{\rm P}_J- {^3{\rm P}}_{J^\prime}$         & \citet{Tayal+Gupta_1999} &---\\
\ion{S}{3}& $^3{\rm P}_J- {^1{\rm D}}_2$                  & \citet{Tayal+Gupta_1999} &---\\
\ion{Fe}{2}& $^6{\rm D}_J- {^6{\rm D}}_{J^\prime}$         & \citet{Ramsbottom+Hudson+Norrington+Scott_2007}
                                                   &---\\
\ion{Fe}{2}& $^6{\rm D}_J- {^4{\rm F}}_{J^\prime}$         & \citet{Ramsbottom+Hudson+Norrington+Scott_2007}
                                                   &---\\
\ion{Fe}{2}& $^6{\rm D}_J-{^4{\rm D}}_{J^\prime}$         & \citet{Ramsbottom+Hudson+Norrington+Scott_2007}
                                                   &---\\
\hline
\end{tabular}
\end{center}
\end{table}

For the
diffuse interstellar radiation field we use the estimate of
\citet{Mathis+Mezger+Panagia_1983}.
Photoionization rates for this radiation field
were taken from \citet[][Table 13.1]{Draine_2011},
calculated using photoionization cross sections
from \citet{Verner+Yakovlev_1995} and 
\citet{Verner+Ferland+Korista+Yakovlev_1996}.

We include secondary ionization and heating by cosmic rays
following \citet{Dalgarno+McCray_1972}.
Charge exchange between H and O is included in our model,
with rate coefficients from
\citet{Stancil+Schultz+Kimura+etal_1999}.

\subsection{Free-Free vs. H$\alpha$\label{sec:freefree vs Halpha}}

We calculate the free-free emission at 41~GHz, $I_\nu(41\GHz)$, to
compare with observations (DDF09):
\begin{equation}
   j_\nu=5.44\times10^{-41}g_{\rm ff}
   T_4^{-0.5}{\rm e}^{-h\nu/kT} n_e^2\ \rm{
   erg\ cm^3\ s^{-1}\ Hz^{-1}\ sr^{-1}}.
\label{eq:free-free}
\end{equation}
where $n_e$ is the electron number density, and the Gaunt factor
\citep{Hummer_1988} 
is accurately approximated by \citep[][eq.\ 10.9]{Draine_2011}
\begin{equation}
g_{\rm ff} \approx \ln\left[
   \exp\left(5.96-
   \frac{\sqrt{3}}{\pi}\ln(\nu_9 T_4^{-1.5})\right)+{\rm e}\right]
~~~.
\end{equation}
where $\nu_9\equiv \nu/$GHz. The H$\alpha$  emission rate is:
\begin{equation}
   j_{\rm H\alpha}=2.82\times10^{-26}T_4^{-0.942-0.031\ln(T_4)}
n_e n(\rm H^+)\,\rm{erg\ cm^3\ s^{-1}\ sr^{-1}}
~~~.
\label{eq:ha}
\end{equation}
corresponding to an effective rate coefficient
\beq
\alpha_{\rm H\alpha}\equiv \frac{4\pi j_{\rm H\alpha}}{h\nu}
= 1.17\times10^{-13}T_4^{-0.942-0.031\ln(T_4)} \cm^3\s^{-1} ~~.
\eeq
The ratio of free-free to
H$\alpha$ depends on both the temperature and
the fraction of the electrons
contributed by H$^+$:
\begin{eqnarray} \nonumber
\psi(T)&\equiv&\frac{j_\nu(41\GHz)}{j_{\rm H\alpha}/h\nu}
\\ \label{eq:psi(T)}
&\approx&
0.0465 \frac{n_e}{n(\rm H^+)}
T_4^{0.442+0.031\ln T_4}
\ln\left[\exp\left(3.913+\frac{3\sqrt{3}}{2\pi}\ln T_4\right)+{\rm e}\right]
\frac{\rm kJy~sr^{-1}}{\rm R}~~.~~~
\label{eq:psi}
\end{eqnarray}
Free-free
emission comes from all ions, therefore in low-ionization gas we have
free-free emission from ions such as C$^++e^-$, whereas H$\alpha$ comes only
from H$^{+}+e^-$.
Therefore, if H becomes almost neutral, $\psi$ is sensitive to the 
gas-phase abundance of elements that can be
ionized by $h\nu<13.6$~eV photons.
\begin{figure}[tb]
\vspace*{-0.3cm}
\begin{center}
\epsscale{0.65} \plotone{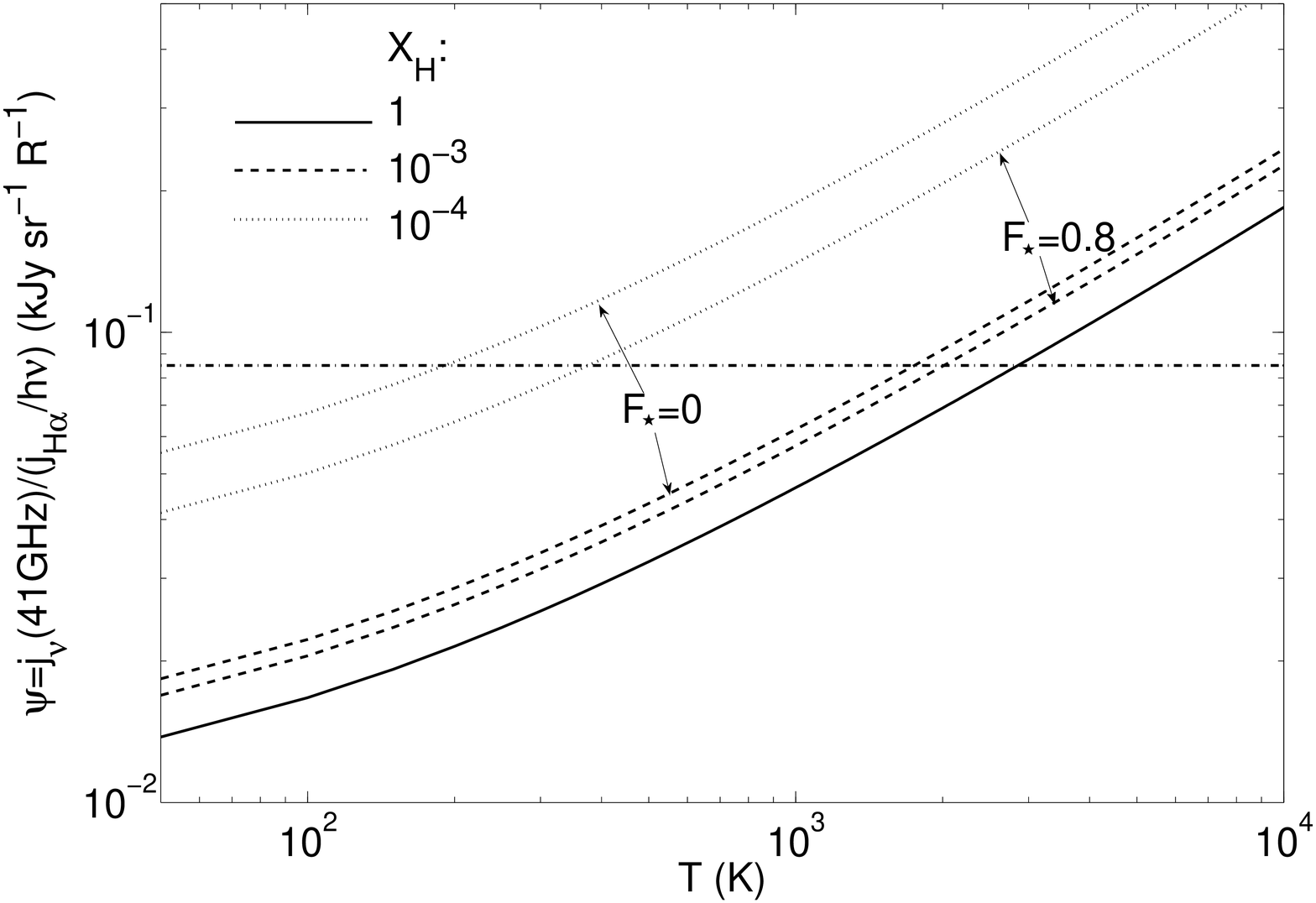}
\vspace*{-0.5cm}
\end{center}
\figcaption{\scriptsize The ratio of $j_\nu(41\rm GHz)/(j_{\rm
    H\alpha}/h\nu)$ as a function of gas temperature $T$ for H
  ionization fraction $x_{\rm H}=1$ (solid curve), $10^{-3}$ (dash
  curves), and $10^{-4}$ (dot curves). For each $x_{\rm H}$, upper
  curves employ gas
  phase elemental abundances for $F_\star=0$ to simulate WIM, lower
  curves employ gas phase elemental abundances for
  $F_\star=0.8$ to simulate CNM. Observed nearly all-sky averaged
  free-free/H$\alpha$ ratio $\sim0.85\,{\rm kJy\,sr^{-1}\,R^{-1}}$
  from {\em WMAP} (DDF09) is indicated by the horizontal dash-dot
  line, corresponding to gas temperature $\sim$3000\,K for fully
  ionized gas.
\label{ffha_T}}
\end{figure}
Figure~\ref{ffha_T} shows
$\psi(T)$
as a function of gas
temperature from $50-10^4$~K, for three different H ionization
fractions $x_{\rm H}\equiv n(\rm H^+)/n_{\rm H}$: $x_{\rm H}=1$ (solid curve), 
$10^{-3}$ (dash curves) and $10^{-4}$
(dot curves), and for two gas phase elemental abundances: $F_\star=0$, representing WIM, and $F_\star=0.8$,
representing CNM. The horizontal dash-dot line indicates the observed 
free-free/H$\alpha$ ratio $\sim0.085\,{\rm kJy\,sr^{-1}\,R^{-1}}$ 
determined from the {\em WMAP} 5 year data (DDF09).
The observed ratio corresponds
to gas temperature $\sim$3000\,K for fully ionized hydrogen.

In addition to emission from the WIM, there will be some H$\alpha$ and free-free emission from \hi~clouds.
The \ion{H}{1} gas in the interstellar medium is found at a wide range of
temperatures.
The majority is in the CNM phase,
at $T\approx10^2\K$, with $n_e/n(\rm H^+)\approx 2$ 
(if $\zeta_{\rm CR}/n(\rm H) \sim 2\times10^{-16}$cm$^{3}$s$^{-1}$,
then $\sim$50\%
of the free electrons are from C$^+$, S$^+$, and other metal ions.)
From this gas we expect 
$\psi\approx 0.017[n_e/n(\rm H^+)]
\kJy\sr^{-1}\,{\rm R}^{-1} \approx 0.034\kJy\sr^{-1}\,{\rm R}^{-1}$.
In addition, a substantial fraction of the \ion{H}{1} is in the WNM phase,
with $T\approx10^3\K$; 
this gas will have
$\psi\approx0.047
\kJy\sr^{-1}\,{\rm R}^{-1}$.
Overall, we estimate that the H$\alpha$ and free-free emission from the \ion{H}{1}
phase will have $\psi_{\rm H\,I}\approx 0.04$.

\subsection{Initial Conditions}

Gas-phase elemental abundances 
are based on the recent study by \citet{Jenkins_2009}. For our standard
model, we use abundances from the model of \citet{Jenkins_2009} with
$F_\star=0$, representing a relatively low level of depletion (see
Section~\ref{sec:discussion} for a discussion of the choice of
$F_\star$). For Ne and Ar (not covered in Jenkins'
study), we assume solar abundances from
\citet{Asplund+Grevesse+Sauval+Scott_2009}.
He/H=0.1 is also assumed. 
In addition, we construct a model with $F_\star=0.25$ (reduced gas
phase abundance for elements that deplete) to explore
the sensitivity to coolant abundances. 
Carbon requires
special attention, since recent work \citep{Sofia+Parvathi_2010}
indicates that the oscillator strength of \ion{C}{2}]2325\AA\ is
larger than previously estimated, implying that gas
phase C abundances estimated from measurements of \ion{C}{2}]2325\AA\ 
could be lowered by a factor $\sim$2.
For this reason, in addition to the standard model
and the $F_\star=0.25$ model, we consider the ``Reduced C'' model in which C
has an abundance two thirds of its $F_\star=0$ value and other
elements all have their $F_\star=0$ values. 

For our standard model we take $n_{\rm H}=0.5\cm^{-3}$ 
and initial temperature $T_i=8000$~K. Under these conditions
the thermal pressure of the photoionized state,
$p/k\approx2.15 n_{\rm H}T\approx 9\times10^3\cm^{-3}\K$, is
comparable to, although somewhat higher than, 
current estimates for pressures in the diffuse ISM.
In our standard model, the initial ionization
fractions (IIF) of all the elements (Table~\ref{table:iif}) are
adopted from \citet{Sembach+Howk+Ryans+Keenan_2000}, where we use
the value in their standard model with parameter $\chi_{\rm{edge}}=0.1$
for fully ionized gas \citep[Table 5]{Sembach+Howk+Ryans+Keenan_2000}, 
corresponding to a fairly soft ionization field. 
Two other IIFs (Table~\ref{table:iif}) considered include values 
from the Orion nebula
\citep{Baldwin+Ferland+Martin+etal_1991}, which corresponds to a
hard radiation field, and a set
of observed values based on various previous studies of the WIM
\citep{
Haffner+Reynolds+Tufte_1999,
Reynolds_2004,
Madsen+Reynolds+Haffner_2006,
Haffner+Dettmar+Beckman+etal_2009}.
The effects of initial gas density and temperature are studied as well. Table~\ref{table:models}
summarizes all the models in our simulations.

The cosmic ray primary ionization
rate for an H atom, $\zeta_{\rm CR}$, is controversial. 
Direct measurement of the cosmic ray flux
\citep{Wang+Seo+Anraku+etal_2002} at $E\ltsim 1$\,GeV is affected by the
solar wind. 
\citet{Webber+Yushak_1983} attempted to correct for solar wind modulation.
Smoothly extrapolating the \citet{Webber+Yushak_1983} spectrum
to lower energy suggests
$\zeta_{\rm CR}\sim1.3\times10^{-17}$~s$^{-1}$. 
However, studies of the ionization conditions inside molecular clouds
\citep{
Black+vanDishoeck_1991,Lepp_1992,
McCall+Huneycutt+Saykally+etal_2003,
Indriolo+Geballe+Oka+McCall_2007} indicate primary ionization
rates $\zeta_{\rm CR}\sim0.3-3\times10^{-16}$~s$^{-1}$,
and we will consider values of $\zeta_{\rm CR}$ within this range.
We adopt $\zeta_{\rm CR}=1\times10^{-16}$~s$^{-1}$ 
for the standard model, and will then explore
varying $\zeta_{\rm CR}$ from $2\times10^{-17}$~s$^{-1}$ to
$5\times10^{-16}$~s$^{-1}$.

\section{Results}\label{sec:results}

\begin{figure}[tb]
\begin{center}
\vspace*{-0.3cm}
\epsscale{0.5} \plotone{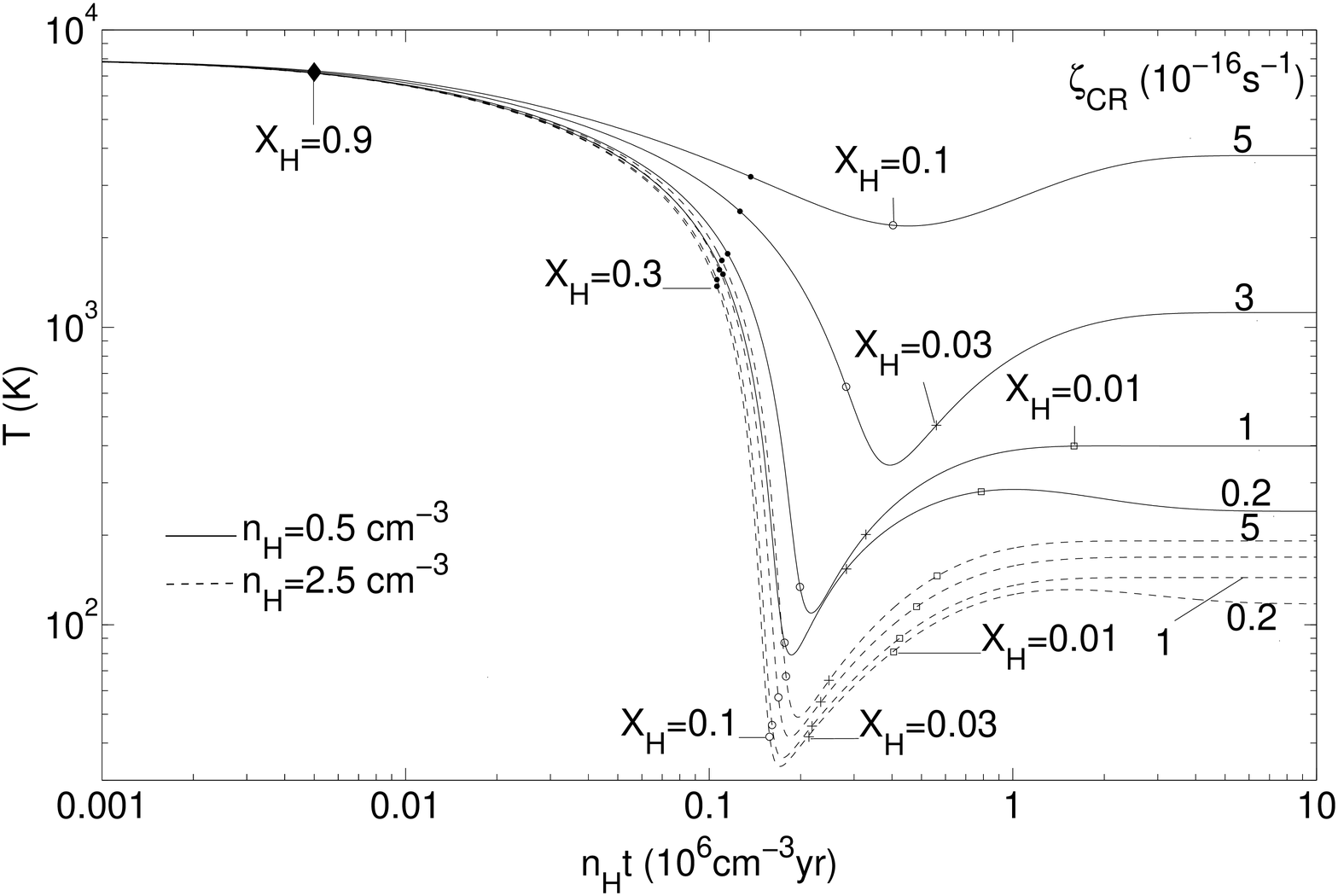} \plotone{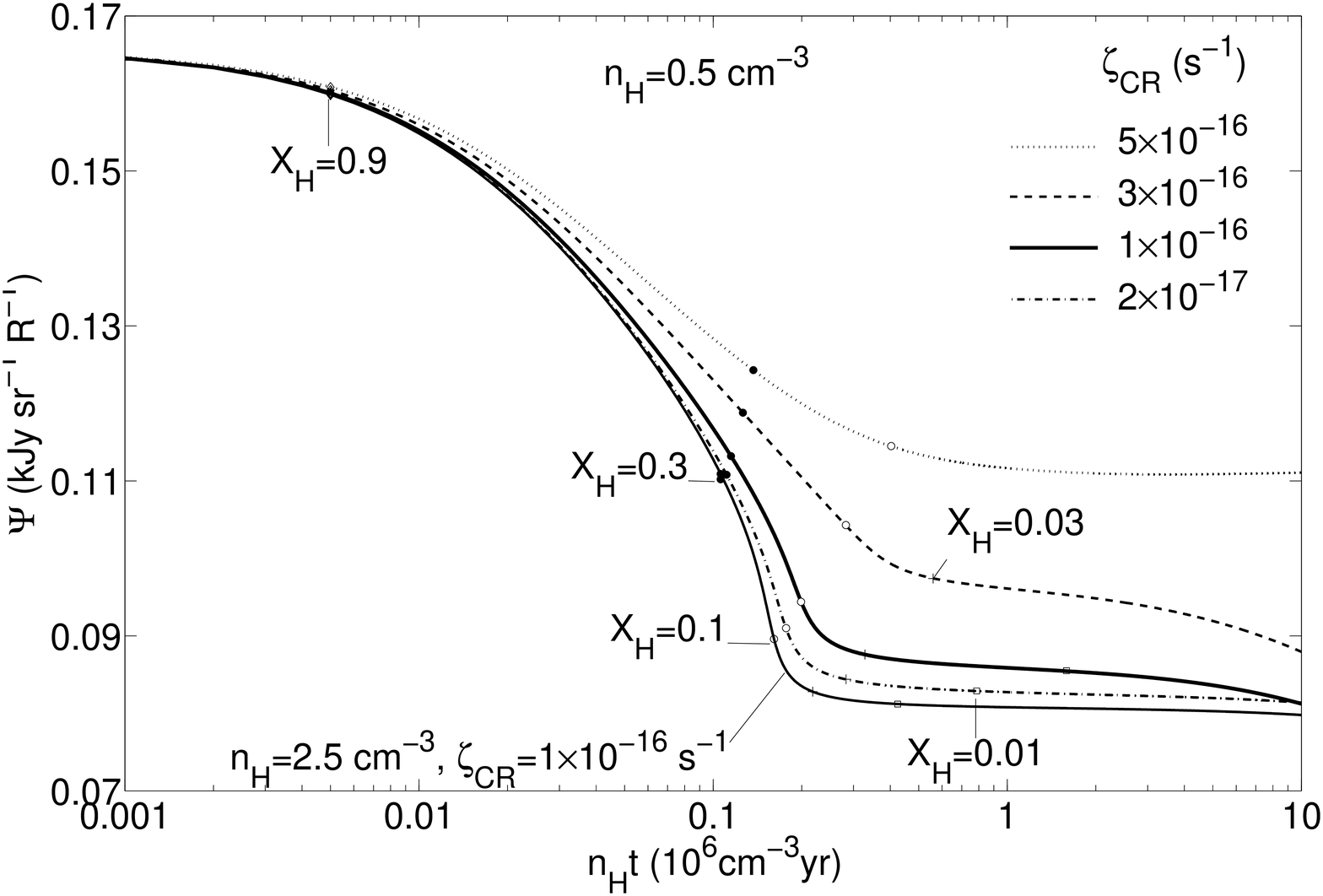}
\vspace*{-0.5cm}
\end{center}
\figcaption{\footnotesize Effect of different density and cosmic ray ionization
  rate. All cases have $g=1/3$ and $F_\star=0$. Top panel: The
  temperature evolution for $n_{\rm H}=0.5$~cm$^{-3}$ (solid curves)
  and $n_{\rm H}=2.5$~cm$^{-3}$ (dashed curves), and different cosmic
  ray ionization rate $\zeta_{\rm CR}$. Bottom panel: The ratio of
  cumulative free-free emission at 41 GHz to cumulative H$\alpha$
  emission for models with different $\zeta_{\rm CR}$ for $n_{\rm
    H}=0.5$~cm$^{-3}$.  Values of $x_{\rm H}=n(\rm H^+)/n_{\rm H}$ are
  given at several points along the curves by different tick marks.
  Note that when $\zeta_{\rm CR}/n_{\rm
    H}>6\times10^{-16}$~cm$^{3}$s$^{-1}$, the final temperature
  $T_f>10^3$K, and $\Phi(n_{\rm H} t=10^7\,{\rm cm}^{-3}\,{\rm yr})>0.09
  {\rm\, kJy\,sr^{-1}\,R^{-1}}$. \label{zetanh}}
\end{figure}

The temperature as a function of time for our standard model is shown
in the top panel of Figure~\ref{zetanh}.
Initially the gas cools
rapidly, reaches a minimum temperature $\sim$100\,K within
$~0.4$~Myr, and then warms up to $400$\,K, approaching its asymptotic
state in several Myr. 
The rise in temperature at $\nH t\gtsim 2\times10^5\cm^{-3}\yr$
occurs because the [C\,II]158$\micron$ cooling declines as
free electrons recombine.  Similar late time reheating has been seen previously
\citep[e.g.,][]{Draine_1978}.
Define the cumulative emission ratio as:
\begin{equation}
   \Psi(t) \equiv 
\frac{\int^t_0 j_\nu(41\GHz) dt^\prime}
     {\int^t_0 (j_{{\rm H}\alpha}/h\nu)dt^\prime}
~~~;
\label{eq:Psi}
\end{equation}
$\Psi$ starts from $\sim$0.17~kJy sr$^{-1}$ R$^{-1}$, corresponding to
the initial temperature $T_i=8000\K$, then drops as the gas cools, to
$\sim$0.08 for 
$10^6$cm$^{-3}\,$yr$\,\ltsim n_{\rm H}t\ltsim10^7$cm$^{-3}\,$yr 
as shown in the bottom panel of
Figure~\ref{zetanh}.
This low value of $\Psi$
is consistent with the observed value of 
$\sim0.085\kJy\sr^{-1}\,{\rm R}^{-1}$ (DDF09).

Altering the parameters in the standard model will change the cooling
history, and we explore the influence of various factors. Based on
their impact, we divide the influential factors into two classes:
major factors, which includes $\zeta_{\rm CR}$, $n_{\rm H}$, abundance of
metal elements and grain depletion, and minor factors: the initial
temperature and IIF.

\subsection{Major factors}

Figure~\ref{zetanh} shows the sensitivity to $\zeta_{\rm CR}$ and
$n_{\rm H}$. For fixed $n_{\rm H}$, the final asymptotic temperature
($T_f$) increases with increasing cosmic ray ionization rate, while
for fixed $\zeta_{\rm CR}$, $T_f$ decreases with increasing $n_{\rm H}$. 
At a higher density the final temperature and free-free to H$\alpha$ emission ratio
become less sensitive to $\zeta_{\rm CR}$. 
For $\zeta_{\rm CR}$ ranging from $2\times10^{-17}$~s$^{-1}$ to
$5\times10^{-16}$~s$^{-1}$, $T_f$ increases from $\sim$240~K to
$\sim$3800~K when $n_{\rm H}=0.5$~cm$^{-3}$, but only from ~120 K to
190 K when $n_{\rm H}=2.5$~cm$^{-3}$ (as shown in top panel of
Figure~\ref{zetanh}).
As $\zeta_{\rm CR}$ is varied from $2\times10^{-17}$ to 
$5\times10^{-16}{~\rm s}^{-1}$,
the emission ratio $\Psi(n_{\rm H}t=10^7\cm^{-3}\yr)$ 
changes about $20\%$ for $n_{\rm H}=0.5$~cm$^{-3}$ 
(as shown in bottom panel of Figure~\ref{zetanh}),
but only about $1\%$ for $n_{\rm H}=2.5$~cm$^{-3}$.
For each density, there is a critical value
of $\zeta_{\rm CR}$, above which the gas remains warm, with 
asymptotic temperature $T>3000$K, and 
$\Psi>0.10$.

\begin{figure}[t]
\begin{center}
\vspace*{-0.3cm}
\epsscale{0.55} \plotone{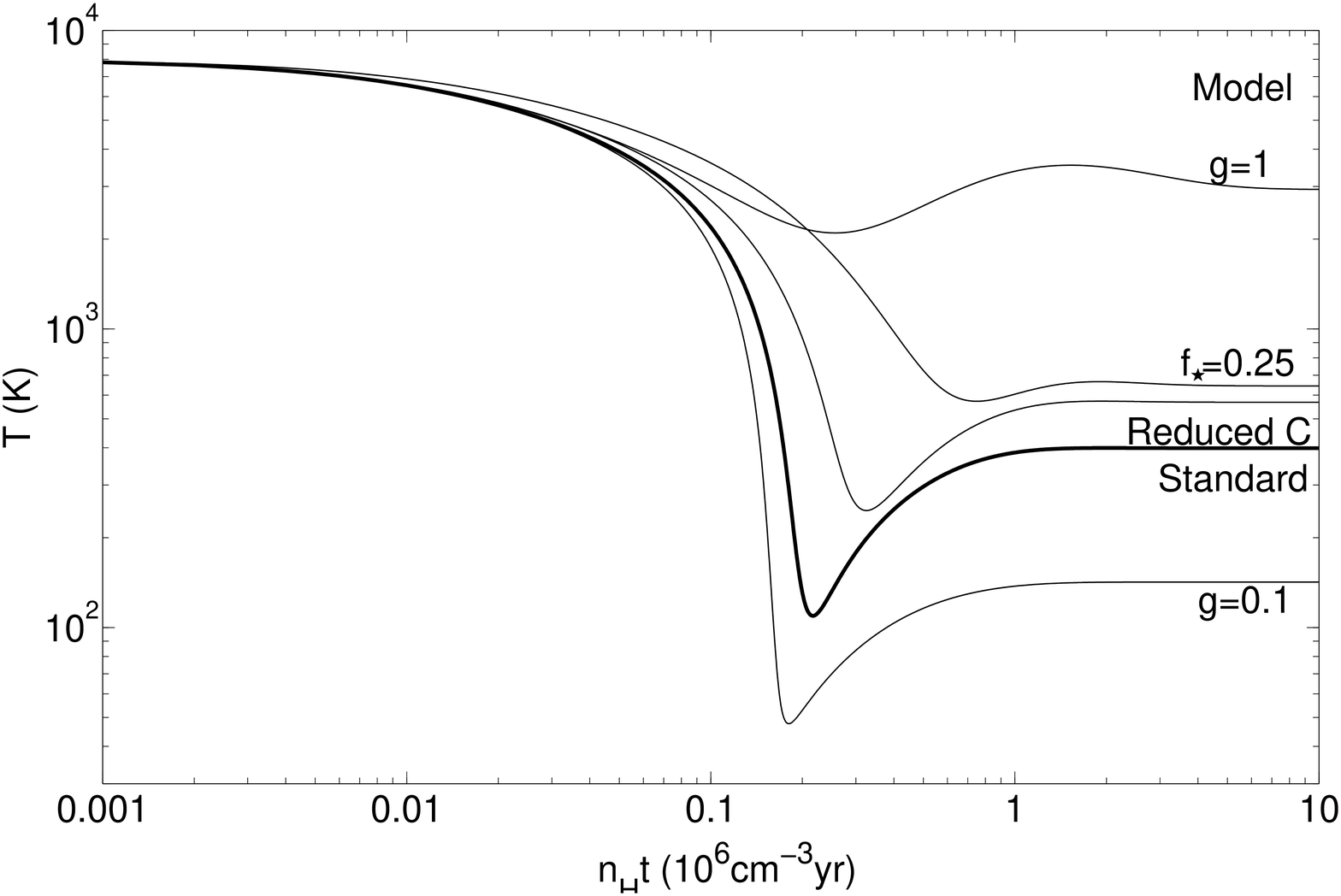} \plotone{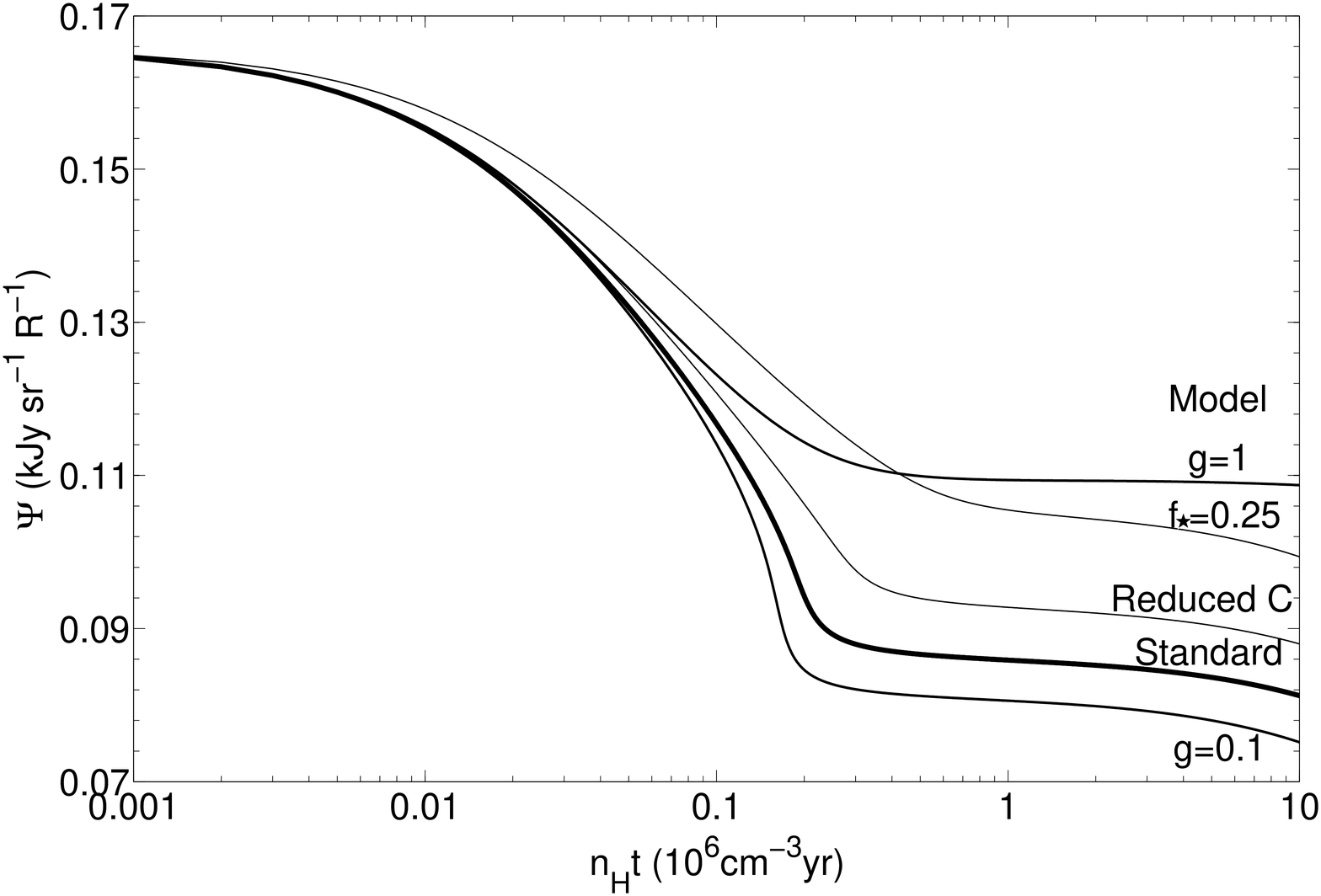}
\vspace*{-0.5cm}
\end{center}
\figcaption{\footnotesize
Effect of varying gas-phase abundances ($F_\star=0.25$ vs.\ $F_\star=0$),
PAH abundance ($g=1$ vs.\ $g=1/3$), 
and C abundance (Reduced C vs.\ $F_\star=0$).
Top panel shows the temperature as a function of time for
different models (indicated by model name), and bottom panel show
the ratio of cumulative free-free emission at 41 GHz to cumulative H$\alpha$
emission as a function of time. \label{major}}
\end{figure}
Elemental abundances affect the cooling process. Figure~\ref{major}
shows the effect of varying the element abundances. In general,
reducing gas phase abundances will reduce the cooling, increase the
final temperature $T_f$, and consequently will raise $\Psi$. As we
discussed in Section~\ref{sec:phy-model}, evidence from
analyzing spinning dust emission in the WIM (DDF09) suggests that the
small polycyclic aromatic hydrocarbons (PAHs) may be underabundant in the WIM
relative to the general ISM. We explore the effect of different grain
reduction factors, with $g=1/3$ as our standard model.
As shown in Figure~\ref{major}, the final value of $\Psi$ is
quite sensitive to the value of the PAH reduction factor $g$.
When $g$ is reduced, the photoelectric heating rate drops;
since the grain-assisted recombination rate also drops, more electrons
and ions will be present in the gas phase, leading to increased cooling
through collisionally-excited lines. The
combined result of the two factors is that the ionized gas recombines
more slowly, cools faster, and reaches lower values of $\Psi$ when $g$ is small.

\subsection{Minor factors}

\begin{figure}[tb]
\begin{center}
\vspace*{-0.3cm}
\epsscale{0.55} \plotone{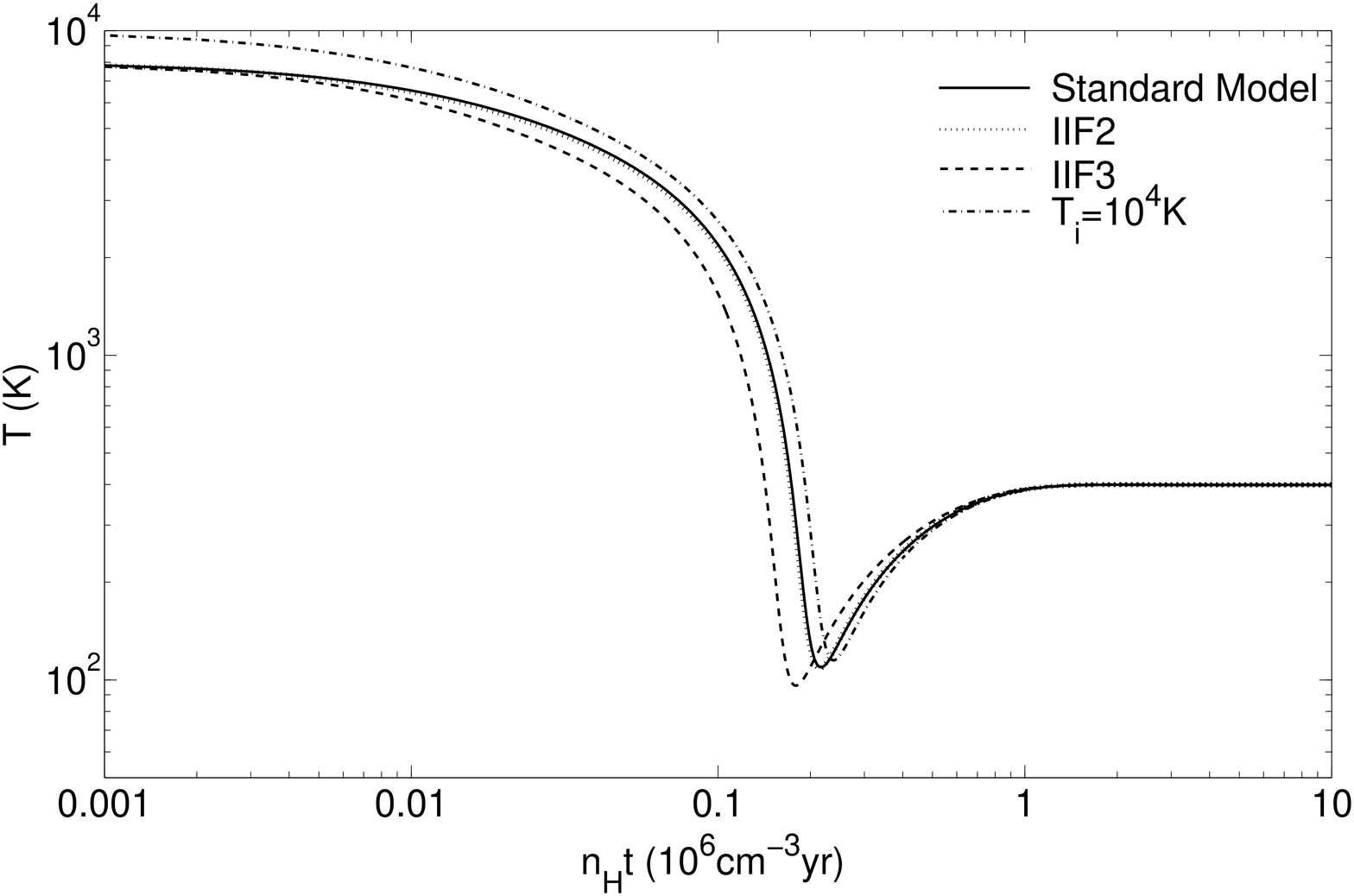} \plotone{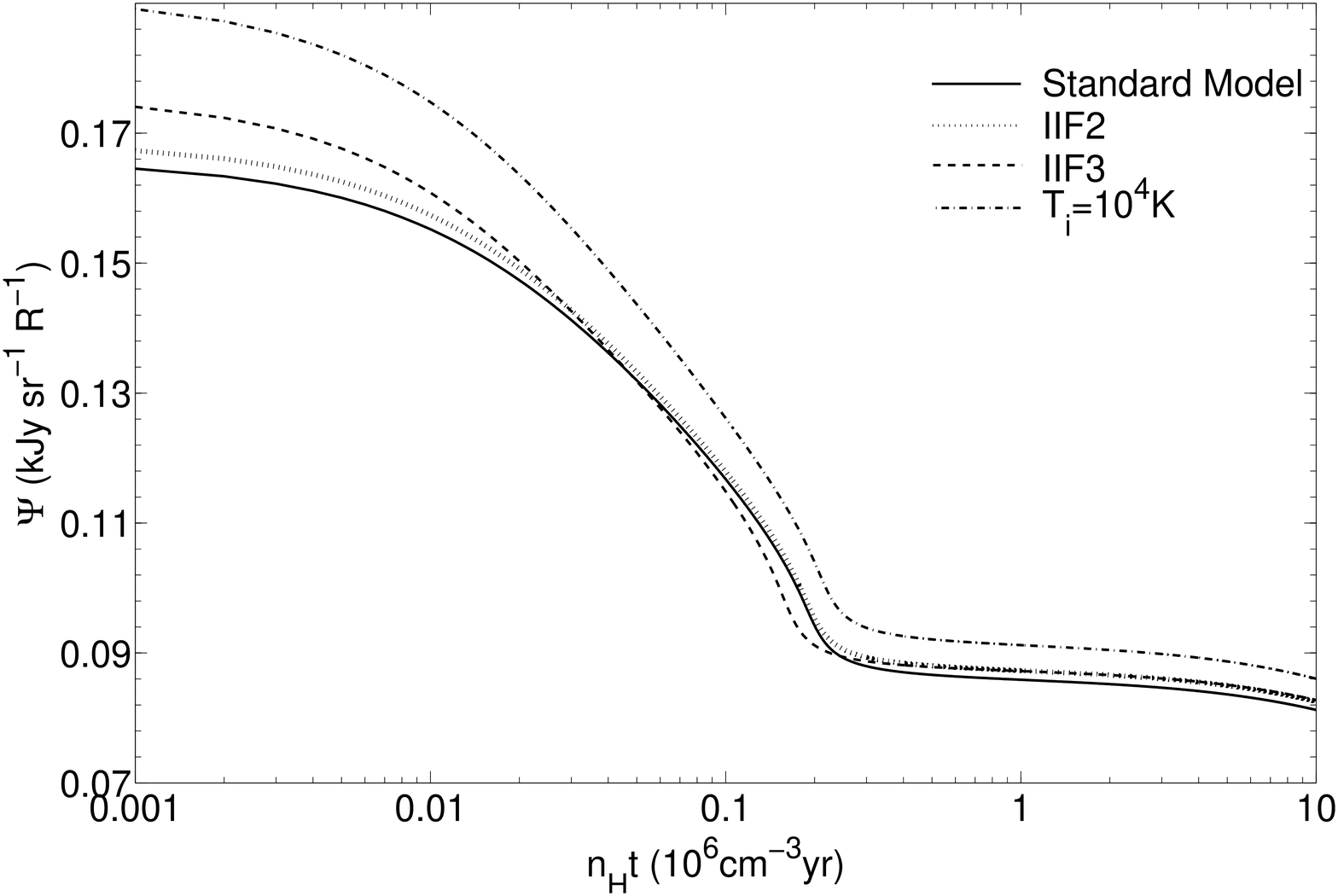}
\vspace*{-0.5cm}
\end{center}
\figcaption{\footnotesize
Effect of different initial conditions (initial
temperature and ionization fractions) on the evolution of
temperature $T$ and the cumulative ratio $\Psi$ of 41\,GHz free-free to
H$\alpha$.
\label{ic}}
\end{figure}

The initial temperature and ionization fractions
also affect the cooling process. 
Quantities like $T$ and $x_{\rm H}$ for models with different initial $T$ and ionization
will evolve differently at the initial stage of the simulations, but
will all converge to the same asymptotic steady-state values.
On the other hand, the cumulative ratio of free-free to H$\alpha$,
$\Psi$,
will differ because it involves integration over time. 
We study the effect of different $T_i$ by
running a model with $T_i=10^4$~K (instead of 8000~K), 
and we study the effect of different IIF by 
running models with IIF2 and IIF3. 
IIF3 assumes Orion Nebula values \citep{Baldwin+Ferland+Martin+etal_1991}, 
formed by a much harder ionization field than the one that forms the WIM, as we
discussed above in Section~\ref{sec:introduction}. 
IIF2 is a set of values based on the literature 
\citep{Madsen+Reynolds+Haffner_2006,
Haffner+Reynolds+Tufte_1999,
Haffner+Dettmar+Beckman+etal_2009,
Reynolds_2004}, 
falling between IIF3 and our standard
model. These three models cover a large range of IIF.
As shown in Figure~\ref{ic}, the results are
insensitive to changes in the
different initial ionization conditions.

\section{A Three Component Model for the Diffuse Emission}
         \label{sec:threegasmodel}

In this section we try to compose a model to simultaneously explain
the observed low free-free to H$\alpha$ emission ratio (corresponding
to $T\sim$~3000 K), and line ratios
such as \nii$\lambda$6583/H$\alpha$ and \nii$\lambda$5755/\nii$\lambda$6583 (indicating $T\gtsim 8000\K$) in the diffuse
emission from the WIM.
We note that the observed low free-free to H$\alpha$
emission ratio (DDF09) comes from measurements which integrate over a large 
fraction of the high-latitude sky, while the measurement of various other
diagnostic lines from the WIM, like \nii$\lambda$6583 or
\nii$\lambda$5755, are limited to relatively high intensity regions.

In our model, 
the observed diffuse emission comes from three components:
\begin{enumerate}
\item \lq\lq hot\rq\rq photoionized gas with $T\approx10^4$K, 
nearly fully ionized, producing 
free-free emission, recombination radiation, and collisionally-excited lines
such as [\ion{N}{2}]$\lambda$6583 and [\ion{N}{2}]$\lambda$5755;
\item  cooling and recombining gas,
which starts from the ``hot'' photoionized phase and cools and
recombines to give H$\alpha$ and free-free emission as well as a small
amount of collisionally-excited lines, such as [\ion{N}{2}]$\lambda$6583;
\item neutral H~I gas, partially in the \lq\lq cold neutral medium\rq\rq (CNM), 
with $T\approx10^2$K, and partially in the ``warm neutral medium'' (WNM),
with $T\approx 10^3\K$; together these emit a small
amount of H$\alpha$ and free-free emission
due to cosmic ray ionization, but negligible metal line emission in the optical.
\end{enumerate}
In addition to the actual emission from the H~I gas, there will also be
reflected light -- both H$\alpha$ and metal lines such
as [\ion{N}{2}]$\lambda$6583 -- that was originally emitted elsewhere, mainly
regular \ion{H}{2} regions, and then reflected by dust grains present in the \hi~gas.
Let $\frefl$ be the fraction of the observed H$\alpha$ that is scattered
light. \citet{Wood+Reynolds_1999} estimated that $\frefl\approx 0.05-0.20$ at
high galactic latitudes. In our calculation, we assume the reflection
coefficients of~[\ion{N}{2}]$\lambda$6583, [\ion{N}{2}]$\lambda$5755
and [\ion{S}{2}]$\lambda$6716 are the same as H$\alpha$,
because the wavelengths are close to H$\alpha$.

The observed intensities are weighted
averages over the three components.
Here we ask what weighting factors are needed to reproduce the 
observed low ratio of free-free/H$\alpha$, as well as other line ratios.

The relative contributions of the three components can be determined from
the observed emission ratios. Among all the
diagnostic line ratios in the WIM, \nii$\lambda$6583/H$\alpha$ and
\sii$\lambda$6716/H$\alpha$ are the two best-studied cases, while
other lines have been studied only in a few select directions
\citep{Haffner+Dettmar+Beckman+etal_2009}. These two line
ratios depend on temperature and ionization fraction of N and S in the gas. 
The high second ionization potential of N (29.60~eV)
protects it from being doubly ionized in the WIM, which has a
generally soft ionization field
\citep{Madsen+Reynolds+Haffner_2006}. This factor along with the
similar first ionization potentials of N and H (14.53 and 13.60~eV)
make N$^+$/N close to
unity in the~\hii~gas 
\citep{
Madsen+Reynolds+Haffner_2006,
Haffner+Reynolds+Tufte_1999}.
On the other hand, some of the S in \ion{H}{2}~regions will be
doubly ionized
due to its low second ionization potential (23.33~eV), 
which leads to uncertainty in 
the predicted
\sii$\lambda$6716/H$\alpha$,
reducing the utility
of \sii$\lambda$6716/H$\alpha$ as a constraint. Nevertheless, we
have calculated the integrated value of \sii$\lambda$6716/H$\alpha$
for the cooling gas (see Table~\ref{table:models}).

Let $\fhot$ and $\fHI$ be, respectively,
the fraction of the observed diffuse H$\alpha$ emitted from the
photoionized gas and H~I clouds, and let $\frefl$ be the
fraction of the observed H$\alpha$ that is actually reflected from dust in \hi.
Then 
$\fcooling=
1-\fhot-\fHI-\frefl$
is the fractional contribution of the cooling material.  The emission
from this three component model depends on four parameters:
$\fhot$, $\fHI$, $\frefl$, and the
temperature $T_{\rm hot}$ of the photoionized component.
For an adopted value of $\frefl$, a physical solution must have 
$0\leq \fhot    \leq1-\frefl$,
$0\leq \fcooling\leq1-\frefl$, and
$0\leq \fHI   \leq1-\frefl$.
We take the observed free-free/H$\alpha$
and \nii$\lambda$6583/H$\alpha$ emission ratios as two constraints to
solve for the relative fraction of the three components.

The instantaneous \nii$\lambda$6583/H$\alpha$ line ratio is:
\beq
   \phi(T)\equiv\frac{j_{\rm [NII]\lambda6583}}{j_{\rm H\alpha}}
   \approx12.4\
   T_4^{0.495+0.040\ln{T_4}}\ {\rm e}^{-2.204/T_4} 
   \frac{n({\rm N^+})/n(\rm H^+)}{7.41\times10^{-5}}
   ~~~.
\label{eq:nii/ha}
\eeq 
where we use the collision strength
$\Omega(^3P_0,^1D_2)=0.303 T_4^{0.0528+0.009\ln{T_4}}$ from
\citet{Hudson+Bell_2005}. The cumulative ratio of [NII]6583 to
H$\alpha$ for cooling gas is 
\beq \Phi(t) \equiv \frac{\int_0^t j_{\rm
    [NII]6583} \,dt^\prime} {\int_0^t j_{\rm H\alpha} \,dt^\prime} =
\frac{\int_0^t \phi(t^\prime)\, j_{\rm H\alpha} \,dt^\prime} {\int
  j_{\rm H\alpha} \,dt^\prime} ~~~.
\label{eq:phi}
\eeq
Cooling gas in our standard model gives $\Phi\approx0.06$ 
(see Table \ref{table:models}).
The WNM and CNM contribute negligible \nii$\lambda$6583.

We take the average value of [\ion{N}{2}]6583/H$\alpha=0.4$, 
based on observations \citep{Reynolds_2004,
Reynolds+Sterling+Haffner+Tufte_2001, Madsen+Reynolds+Haffner_2006}.
Then we must have
\begin{equation}\label{eq:NII/Halpha}
0.4
= (\fhot+\frefl) \phi(T_{\rm hot}) + (1-\fhot-\fHI-\frefl)\Phi
~~~.
\end{equation}
Note that in this equation, we assume the reflected component
has the same [\ion{N}{2}]$\lambda$6583/H$\alpha$ ratio as the
hot ionized component, as discussed previously.

The ratio of
all-sky free-free to H$\alpha$ emission is 
$I_\nu(41\GHz)/I({\rm H}\alpha) \sim0.085\rm{~kJy~sr^{-1}~R^{-1}}$ (DDF09).
Thus we must have
\beq \label{eq:freefree/Halpha}
0.085\kJy\sr^{-1}\,{\rm R}^{-1} =
\fhot \psi(T_{\rm hot})
+
(1-\fhot-\fHI-\frefl)\Psi
+
\fHI\psi_{\rm H\,I}
~~~.~~~
\eeq
For trial values of $T_{\rm hot}$ and $\frefl\geq0$, we can use the observed
$I([{\rm N\,II}]6583)/I({\rm H}\alpha)$
and
$I_\nu(41{\rm GHz})/I({\rm H}\alpha)$
to determine $\fhot$ and
$\fHI$, by solving the two linear equations
(\ref{eq:NII/Halpha}) and (\ref{eq:freefree/Halpha}).
The coefficients $\psi(T_{\rm hot})$ and $\phi(T_{\rm hot})$ are obtained from
eq.\ (\ref{eq:psi(T)}) and (\ref{eq:nii/ha}), while
$\Psi\approx0.08$ and $\Phi\approx0.06$ are obtained from the appropriate
simulations as their asymptotic values (see Table~\ref{table:models} 
as well as Figure~\ref{zetanh}).
Based on the discussion in \S\ref{sec:freefree vs Halpha}, we
take $\psi_{\rm H\,I}\approx 0.04$.
Physical solutions must have
$0\leq\fhot\leq1$, $0\leq\fHI\leq1$, and $0\leq(1-\fhot-\fHI-\frefl)$.
\begin{figure}[tb]
\begin{center}
\vspace*{-0.3cm}
\epsscale{0.7}
\plotone{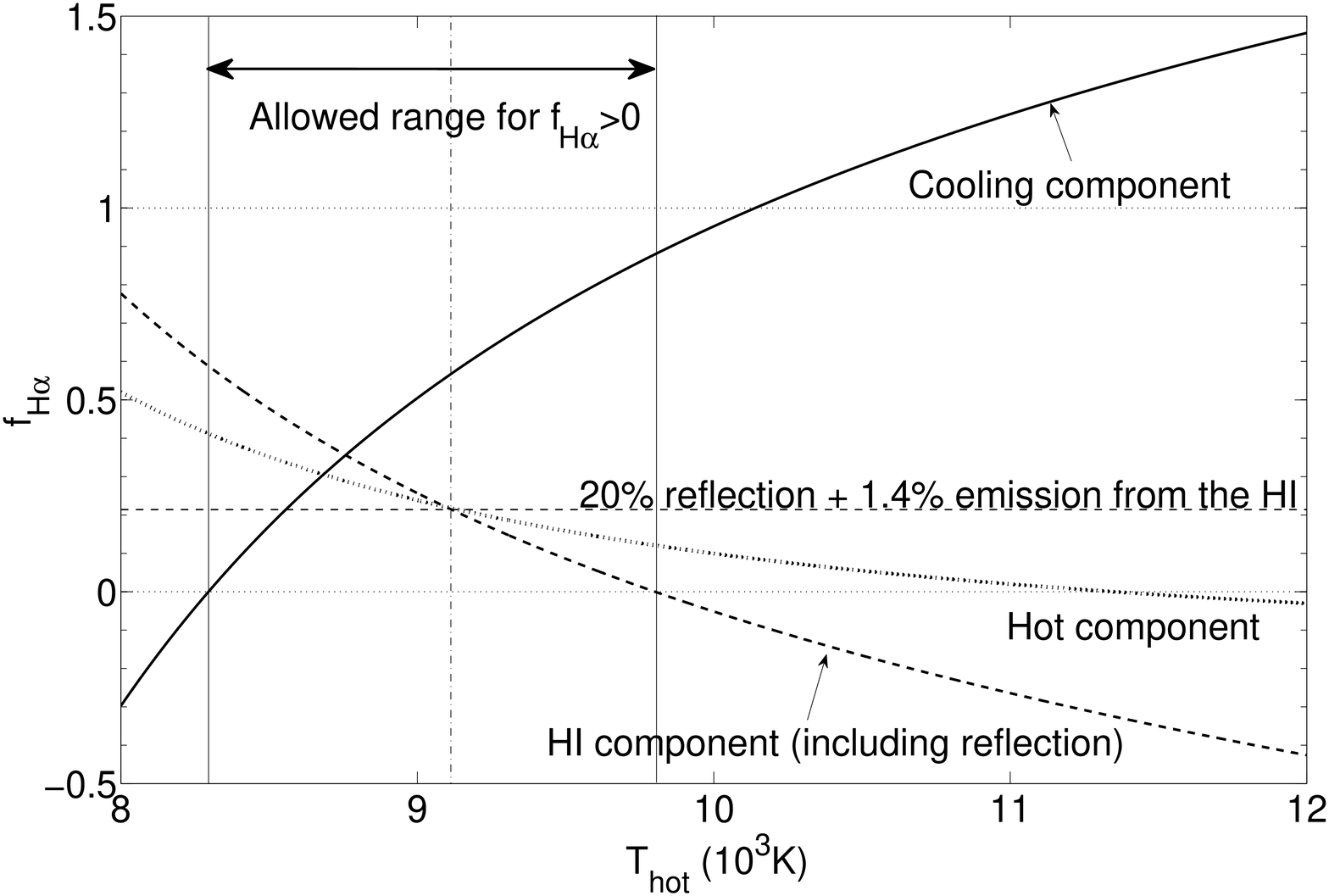}
\vspace*{-0.5cm}
\end{center}
\figcaption{\footnotesize H$\alpha$ fractions for the three gas
  components based on two constraints ($I_\nu(41\,{\rm GHz})/{\rm H}\alpha
    =0.85\,{\rm kJy\,sr^{-1}R^{-1}}$ and [N\,II]$\lambda$6583/H$\alpha$=0.4),
  as a function of hot (photoionized) gas temperature $T_{\rm
    hot}$. The equations are solved under these assumptions:
  $\frefl=0.20$, 
  $\fHI=0.014$, $n({\rm H^+})/n_e=0.5$ in the H\,I (corresponding to
  $\zeta_{\rm CR}\sim 1\times10^{-16}$s$^{-1}$ in CNM), and
  He$^+$/He=0.3 in the hot ionized gas, as in IIF1.
  The fact that $\frefl+\fHI\approx\fhot$ is coincidental.
\label{refl0.2}}
\end{figure}
Figure~\ref{refl0.2} shows one example, 
where the reflected H$\alpha$ fraction is set to $20\%$,
$n({\rm H^+})/n_e=0.5$ in the \hi~phase and $n({\rm He^+})/n(\rm He)=0.3$ in
the hot gas phase. 
The sensitivity to the first two parameters will be discussed below.

In addition to the requirement that the fractions of all the three
components have to be positive, which,
for $\frefl\approx 0.2$, requires 8300$\leq T_{\rm
  hot}\leq 9800$~K (see Figure~\ref{refl0.2}), there is a second
restriction coming from the fact that the total amount of H$\alpha$
emission from the \hi~phase is determined by the cosmic ray ionization
rate. Consider a uniform layer of cold neutral gas with a column
density $N(\rm HI)$, with cosmic ray ionization of H balanced by
case B recombination as well as grain assisted recombination; the
H$\alpha$ intensity from this component at latitude {\it b} is
\begin{equation}
\frac{I_{\rm H\alpha}}{h\nu}=
\frac{1}{4\pi}\frac{3\times10^{20}\cm^{-2}}{|\sin b|}\zeta_{\rm CR}(1+\phi_{\rm CR})\frac{\alpha_{\rm H\alpha}/\alpha_{\rm B}}{1+\alpha_{\rm gr}/x_e\alpha_{\rm B}}~~~,
\label{eq:coolgasha}
\end{equation}
where $x_e\equiv n_e/n_{\rm H}$, $\phi_{\rm CR}\sim0.67$ is the number
of secondary ionization per primary ionization in neutral gas
\citep{Dalgarno+McCray_1972}, and $N{(\rm HI)}\approx
3\times10^{20}|\sin b|^{-1}\cm^{-2}$ \citep{
  Radhakrishnan+Murray+Lockhart+Whittle_1972,
  Dickey+Salpeter+Terzian_1978}.  Thus
\begin{equation}
   \frac{I_{\rm H\alpha}}{h\nu}\approx\frac{0.01}{|\sin b|}\left(\frac{\zeta_{\rm CR}}{{10^{-16} s^{-1}}}\right)\mbox{R}~~~,
\label{eq:coolgasha2}
\end{equation}
where the \ion{H}{1} is taken to be a 1:1 mixture 
of CNM and WNM material.\footnote{We have taken 
$\phi_{\rm CR}\approx0.67$, 
$\alpha_{\rm H\alpha}/\alpha_{\rm B}\approx0.6$, 
$\alpha_{\rm gr}/x_e\alpha_{\rm B}\approx10$ for 
CNM with 
$T\approx 100\K$, 
$n_e\approx 0.01$,
$\nH\approx 30\cm^{-3}$ \citep{Draine_2011};
$\phi_{\rm CR}\approx0.5$, 
$\alpha_{\rm H\alpha}/\alpha_{\rm B}\approx0.5$ and
$\alpha_{\rm gr}/x_e\alpha_{\rm B}\approx0.5$ for $T\sim5000$K WNM;
and $N_{\rm CNM}\approx N_{\rm WNM}\approx \frac{1}{2}N_{\rm HI}$.}

\begin{figure}[tb]
\begin{center}
\vspace*{-0.3cm}
\epsscale{0.50}
\plotone{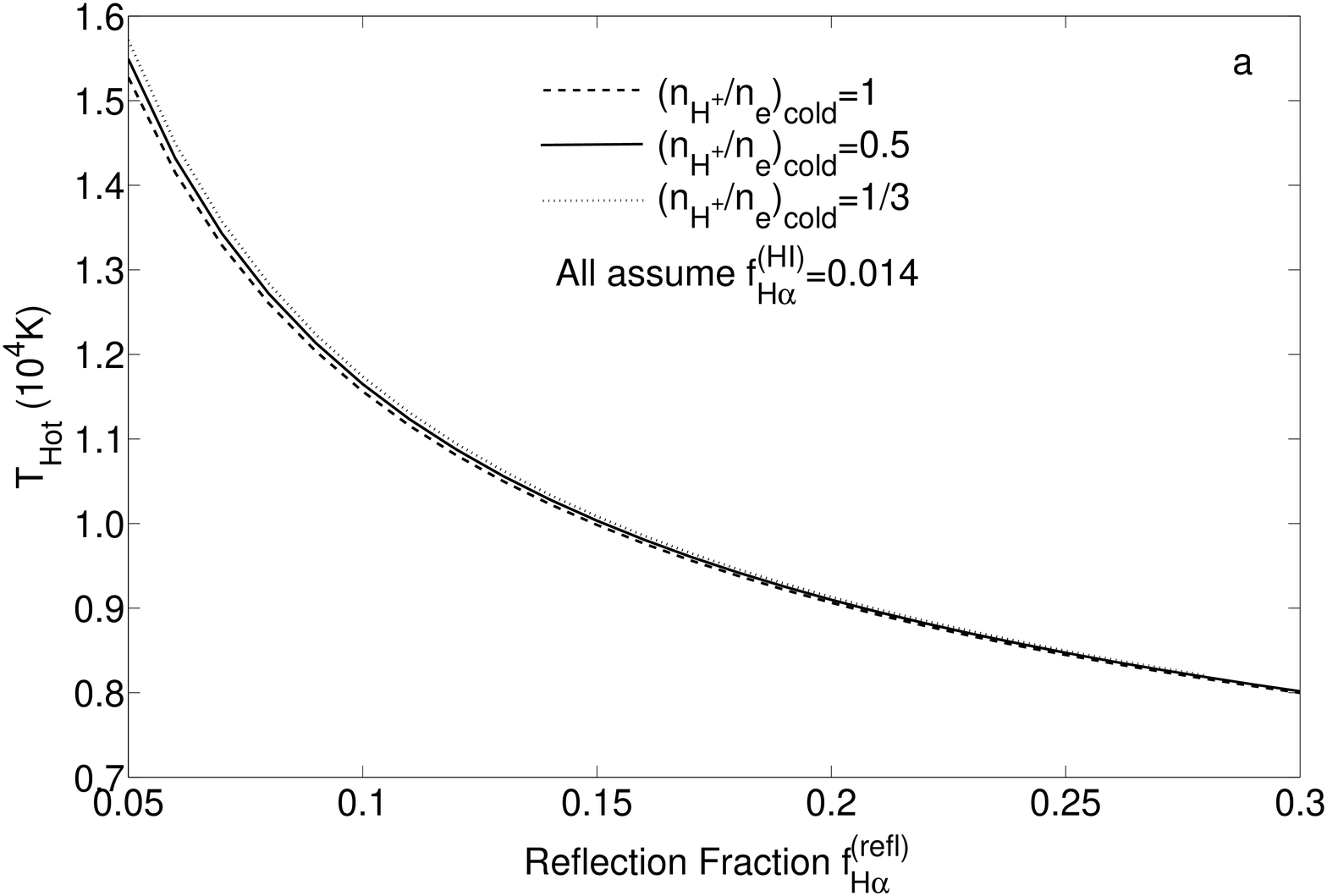}\hspace*{-0.2cm}
\plotone{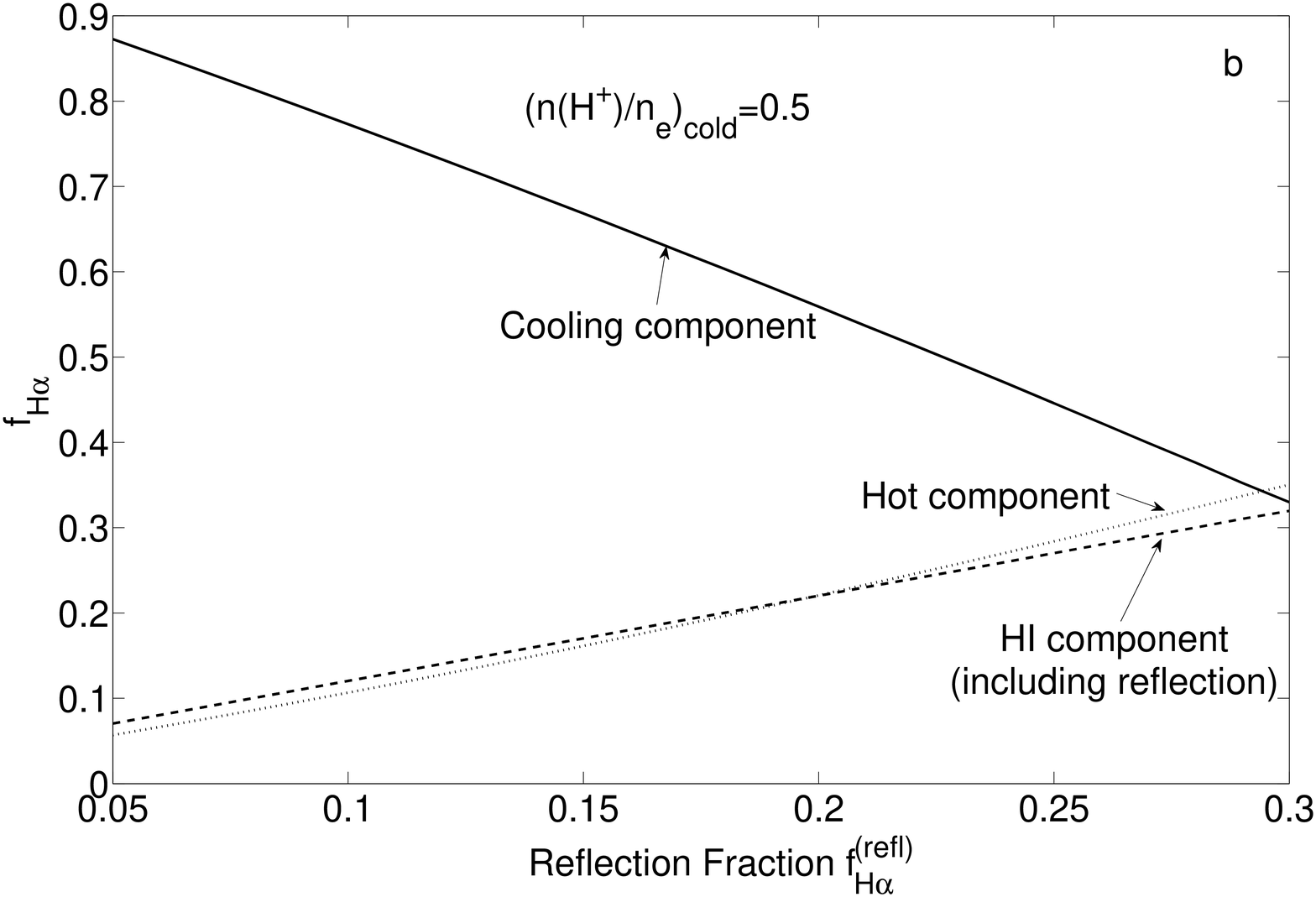}
\vspace*{-0.5cm}
\end{center}
\figcaption{\footnotesize
(a) Temperature $T_{\rm hot}$ and (b) fractional H$\alpha$ contributions
  from different ISM components, as functions of the
  reflected H$\alpha$ fraction $\frefl$,
  based on two constraints ---
  $I_\nu(41\,{\rm GHz})/{\rm H}\alpha = 0.85\,{\rm kJy\,sr^{-1}R^{-1}}$ and
  [N\,II]$\lambda$6583/H$\alpha = 0.4$ --- for $\frefl$ ranging from
  $5\%-30\%$. It is assumed that only
  $\fHI\approx0.014$ of the H$\alpha$ is emitted by the CNM and WNM
  components (see text). Panel (a) also shows the dependence of the
  solution ($T_{\rm hot}$) on the value of $n({\rm H}^+)/n_e$ in the H\,I gas.
  Our standard model assumes $\zeta_{\rm CR}=1\times10^{-16}$s$^{-1}$,
  which corresponds to $n({\rm H}^+)/n_e\approx0.5$ in the ``cold''
  phase, as discussed in Section \ref{sec:threegasmodel}.
\label{solution_vs_refl}}
\end{figure}
The distribution of H$\alpha$  for $|{\it b}|\geq 10^\circ$
has recently been measured by \citet{Hill+Benjamin+Kowal+etal_2008},
who found the full WIM to be fitted on average by 
$I_{\rm H\alpha}/h\nu\approx (0.625\pm0.002)|\sin b|^{-1}$~R.  Comparing
these two results, the fraction of H$\alpha$ {\it emitted} by the 
\ion{H}{1} should only
be $\fHI\approx0.014(\zeta_{\rm CR}/10^{-16}{\rm s}^{-1})$.
In Figure~\ref{solution_vs_refl}
we take $\fHI=0.014$ and consider different values of $\frefl$.
For each $\frefl$, the H$\alpha$/free-free and \nii/H$\alpha$ constraints
serve to determine $\fhot$ and $T_{\rm hot}$, as shown in Figure~\ref{refl0.2}.
For $\frefl$ ranging from $5\%$
to $30\%$,
panel (a) in Figure~\ref{solution_vs_refl} 
gives $T_{\rm hot}$ ranging from 8000--15000\,K.
For $\frefl=0.2$, we find $T_{\rm hot}\approx 9100\K$.

\begin{figure}[t]
\begin{center}
\vspace*{-0.3cm}
\epsscale{0.60} \plotone{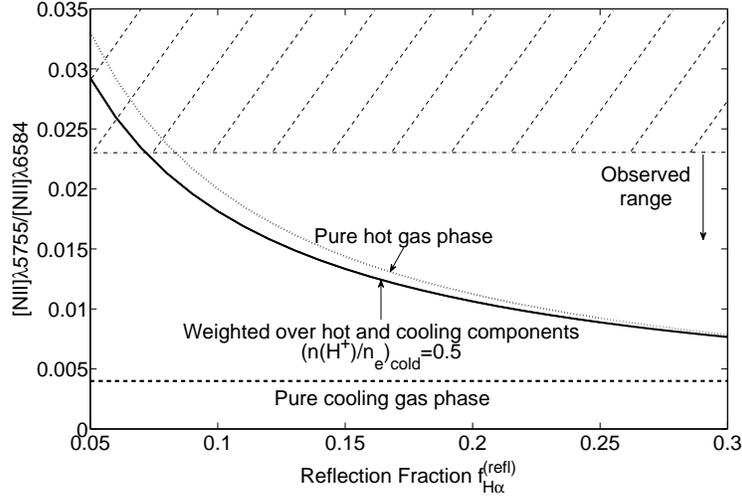}
\vspace*{-0.5cm}
\end{center}
\figcaption{\footnotesize
  [N\,II]$\lambda$5755/[N\,II]$\lambda$6583 averaged over
  components (solid curve) as a function of reflection
  fraction based on the solution in panel (b) of
  Figure~\ref{rnii_vs_refl}. Dashed curve and dotted curves show
  [N\,II]$\lambda$5755/[N\,II]$\lambda$6583 associated with pure
  cooling and pure hot gas phase, giving lower and upper limits of the
  ratio on individual sightlines. The result agrees with 
  observations in \citet{Madsen+Reynolds+Haffner_2006}, although
  comparison between the two is not straightforward (see discussion in
  Section~\ref{sec:threegasmodel}).
\label{rnii_vs_refl}}
\end{figure}
One uncertainty when solving equations (\ref{eq:NII/Halpha}) and
(\ref{eq:freefree/Halpha}) is $n(\rm H^+)/n_e$ in the \hi, which
affects $\psi_{\rm H\,I}$ (equations \ref{eq:psi}).  Assuming
$\zeta_{\rm CR}\sim10^{-16}$s$^{-1}$, $T_{\rm CNM}\sim10^2$~K and
$n_{\rm CNM}\sim30$cm$^{-3}$, about half of the electrons come from
metal elements, so the free-free/H$\alpha$ ratio for the \hi~will be
$\Psi_{\rm HI}\approx 0.04$, as discussed in
Section~\ref{sec:freefree vs Halpha}. We explore the effect of this
uncertainty on $T_{\rm hot}$ in the range $0.02-0.06$, corresponding
to $(n({\rm H^+})/n_e)_{\rm cold}=1$ and $\frac{1}{3}$, as shown in
panel (a) of Figure~\ref{solution_vs_refl}. We find this
uncertainty does not significantly affect the derived solutions.

In addition to \nii$\lambda$6583/H$\alpha$, the
\nii$\lambda$5755/\nii$\lambda$6583 ratio is another indicator of the
temperature in the ionized medium. While \nii$\lambda$6583 is one of
the strongest lines in the WIM, comparable in intensity to H$\alpha$,
\nii$\lambda$5755 is extremely hard to detect due to its low
intensity. Nevertheless, several measurements of this line ratio have
been reported (Table 6 in \citet{Madsen+Reynolds+Haffner_2006}). The
\nii$\lambda$5755/\nii$\lambda$6583 data have a large scatter, ranging
from 0.023 (corresponding to a temperature over 12000~K) down to the
observational upper limit 0.002 (corresponding to a temperature below
6000~K). In our WIM model the \nii$\lambda$5755/\nii$\lambda$6583
ratio is determined by the relative amounts of emission from the hot
and cooling components, with the ratio for the cooling gas about 0.004
(almost model independent, as shown in Table~\ref{table:models}). The
solid curve in Figure~\ref{rnii_vs_refl} shows the weighted
\nii$\lambda$5755/\nii$\lambda$6583 ratio based on the solutions in
the panel (b) of Figure~\ref{solution_vs_refl} for $\frefl$
ranging from 5\%--30\%, under the assumption that the
\nii$\lambda$5755/\nii$\lambda$6583 ratio of the reflected light is
the same as its value in the hot ionized gas, as we discussed
previously. Note that this result is a {\it weighted average}
value over all the components in the WIM --
measurements along individual sightlines will vary. The observed
\nii$\lambda$5755/\nii$\lambda$6583 value along a specific sightline
depends on the weight of the three components along this sightline,
which may differ from the weights that give the observed all sky
averaged free-free to H$\alpha$ and \nii$\lambda$6583 to H$\alpha$
ratio. The dashed curve and the dotted curve in
Figure~\ref{rnii_vs_refl} show \nii$\lambda$5755/\nii$\lambda$6583 of
the cooling gas and the hot gas, which could be interpreted as the
lower and upper limit of the average
\nii$\lambda$5755/\nii$\lambda$6583 value under extreme weighting
cases. Nevertheless, the average weighted
\nii$\lambda$5755/\nii$\lambda$6583 value ranges from 0.008 to 0.025,
in general agreement with the observed values and upper limits in
\citet{Madsen+Reynolds+Haffner_2006}. Individual sightlines could have
values of \nii$\lambda$5755/\nii$\lambda$6583 as low as 0.004, if the
sightline is dominated by the cooling gas.

In sum, if the cooling gas is described by our standard model, the
temperature of the hot gas and the fractions of the three components
could be solved for so that a weighted sum of fractions of the three
components:
\begin{enumerate}
\item reproduces the low observed free-free/H$\alpha$ ratio;
\item reproduces the observed \nii$\lambda$6583/H$\alpha$ ratio
\item has \nii$\lambda$5755/\nii$\lambda$6583 in agreement with
observations;
\item has a small fraction ($\sim$1.4\%) of
the H$\alpha$ {\it emitted} from \hi gas;
\item includes a reflected component accounting for $\sim$20\% of the
  total H$\alpha$
\end{enumerate}

We emphasize again that the three-component model presented in this
section is aimed to explain {\it averaged} observational results;
individual sightlines will generally differ from the average
because (1) the relative weights of its components differ from the
average values,
the temperature $T_{\rm hot}$ and the ionization conditions in the 
\ion{H}{2} component may vary,
and (3) the cooling component along an individual sightline may not
be an average over the entire cooling and recombination.
There is evidence
that the physical conditions in the WIM do vary from sightline to
sightline, and even among different velocity components within one
sightline \citep{Madsen+Reynolds+Haffner_2006}. Different regions in
the WIM will have different densities, cosmic ray ionization rates,
elemental abundances, and histories. The three component model
presented here is highly idealized, but it appears to provide a
physical framework that is consistent with the observations.

\section{Discussion}\label{sec:discussion}

When the $h\nu >13.6$~eV starlight is cut off at the beginning of the simulation, the hot gas
begins to cool and recombine. The temperature, particle density and gas pressure decrease.
In general, the cooling time scale is short, as shown in
Figure~\ref{zetanh}: for $n_{\rm H}=0.5$~cm$^{-3}$, 
$T$ falls below $10^3$~K within 0.3 Myr.
If the photoionized gas was initially overpressured
(relative to its surrounding), it would initially be expanding,
resulting in adiabatic cooling after the photoionizing
source turns off at $t=0$. Conversely, 
if the photoionized gas was in pressure
equilibrium with a confining medium at $t=0$, it would begin to undergo
compression as it cools and recombines at $t>0$.
However, if the $n_{\rm H}=0.5\rm\, cm^{-3}$, photoionized region
is $\sim$5 pc or larger, the cooling time will be short compared to the sound
crossing time, and the effect of adiabatic expansion
or compression should be of secondary importance.
In the present work we assume isochoric evolution, which appears to be
a reasonable approximation during the cooling phase.

In modeling the cooling gas, we assumed case B
recombination. In the real WIM, some of the $h\nu>13.6~$eV 
recombination radiation
might escape from the WIM and get absorbed by nearby cold or warm neutral
gas ($T\leq1000$~K); the resulting ionized H in these cooler media would 
recombine, emitting
H$\alpha$ and free-free emission, with a ratio
of free-free emission to H$\alpha$ appropriate to $T\ltsim1000$~K, lower
than the ratio for the cooling gas. 
This would have the effect of decreasing the ratio of free-free emission to
H$\alpha$ below the value calculated for the cooling gas model,
allowing the observed low free-free/H$\alpha$ ratio to be
explained by a somewhat smaller value of $\fcooling$,
the fraction of the H$\alpha$ contributed by cooling and recombining gas.
However, we expect this ``escape'' of ionizing photons to be minimal: 
the mean free path of $\sim$14\,eV photons in partially recombined
H is short, and we expect the case B on-the-spot treatment
to be a valid approximation.

In our standard model, we choose $F_\star=0$ for the
elemental abundances in the WIM. In Jenkins' model, the nonzero initial
depletion ($F_\star=0$) is identified with the
gas-phase abundances in a warm, low-density medium
\citep{Spitzer_1985, Savage+Sembach_1996}, 
and should therefore apply to the WIM.
Also, as mentioned above, there is evidence (DDF09) showing that the
PAHs in the WIM are underabundant by a factor of $\sim$3. Our
calculation shows that both factors are crucial for cooling the gas quickly
enough and to a low enough temperature to be able to reproduce the
observed low free-free/H$\alpha$ ratio. As shown in
Figure~\ref{major}, if there is no grain depletion or there is
appreciable depletion of coolants (such as for $F_\star=0.25$), the
recombining gas stays relatively hot in the final steady state, and
gives an unacceptable high integrated free-free/H$\alpha$ ratio.

Cosmic rays heat the gas directly, and they also raise the electron density
$n_e$.
If $n_e$ is low, photoelectron emission causes dust grains and PAHs to
become positively charged, reducing the photoelectric heating rate.
Thus increased cosmic ray ionization, by lowering the charge state of the
dust and PAHs, has the effect of increasing
the dust photoelectric heating rate.
For our cooling gas model to be able to reproduce the low
free-free/H$\alpha$ ratios that are observed, the ratio of cosmic ray
primary ionization rate to gas density $\zeta_{\rm CR}/n_{\rm H}$
should not exceed $\sim6\times10^{-16}\cm^3\s^{-1}$.  
For
$\zeta_{\rm CR}/n_{\rm H}\gtsim5\times10^{-16}\cm^3\s^{-1}$ 
(e.g, $n_{\rm H}=0.5\cm^{-3}$
and $\zeta_{\rm CR}=5\times10^{-16}\s^{-1}$, with $\zeta_{\rm
  CR}/n_{\rm H}=1\times10^{-15}\cm^3\s^{-1}$), the cosmic ray
ionization maintains $n({\rm H}^+)/\nH\geq0.1$, and grain
photoelectric heating can sustain the gas at
$T\geq10^3$~K. 
Observations of ${\rm H}_3^+$ give estimates of
$\zeta_{\rm CR}\approx(0.5-3)\times10^{-16}\cm^3\s^{-1}$, with an
average of $\sim2\times10^{-16}\cm^3\s^{-1}$
\citep{Indriolo+Geballe+Oka+McCall_2007}. If $\zeta_{\rm
  CR}\approx2\times10^{-16}\cm^3\s^{-1}$, then the gas density
cannot be much smaller than $n_{\rm H}=0.5\cm^{-3}$ if the gas is to
cool with $\Psi\leq0.085$, as required to explain the low observed
value of free-free/H $\alpha$. This high initial value of
$n_e=0.5\cm^{-3}$ seems at first sight to be at odds with the study by
\citet{Hill+Benjamin+Kowal+etal_2008}, which concluded that the most
probable value of $n_e$ is only $n_e\approx0.03\,$cm$^{-3}$ in turbulent
models of the WIM. Note, however, that in our standard model, the
cooling gas ends up with $x_{\rm H}<0.03$ and
$n_e\approx0.015$cm$^{-3}$.

Because the observed line emission and free-free emission are weighted
sums over 3 components, one of which (the cooling phase) itself has a
range of temperature, there is no way to measure emission ratios
associated with different phases unless the components could be
separated. Temperatures derived from different line ratios need not
agree.
Moreover, even for a single line ratio, the contributions from
different components may vary significantly from sightline to
sightline, or with velocity on a single sightline. This may explain
the large scatter in the physical conditions deduced from different
line ratios.

The present model envisages intermittent
photoionization events in the diffuse high-latitude gas, 
followed by cooling and recombination.
Suppose that the probability per unit time of a photoionization event
is $\tau_{\rm pi}$, with the ionizing radiation 
lasting a time $\tau_{\rm H\,II}$
before the photoionization switches off and the gas begins to cool and
recombine.
During the recombination phase, 
the number of H$\alpha$ photons per recombining H$^+$ is $\sim$0.5.
The ratio of the H$\alpha$ emission from the photoionized gas to the
emission from the cooling gas is
\beq
\frac{\fhot}{\fcooling}\approx 
\frac{\tau_{\rm pi}^{-1}\times n_e \alpha_{{\rm H}\alpha}\tau_{\rm H\,II}}
     {\tau_{\rm pi}^{-1}\times 0.5}
\eeq
Hence,
\beq
\tau_{\rm H\,II}\approx \frac{0.5}{n_e\alpha_{{\rm H}\alpha}}
\frac{\fhot}{\fcooling}
\eeq
If $\frefl\approx 0.2$, then
$\fhot/\fcooling\approx 0.22/0.56\approx 0.39$,
and the duration of the photoionization phase is only
$\tau_{\rm H\,II}\approx 1\times10^5 (0.5\cm^{-3}/n_e)\yr$.

Given that O star lifetimes are $\sim3\times10^6\yr$,
such a short value of $\tau_{\rm H\,II}$ seems surprising.
Longer values of $\tau_{\rm H\,II}$ can be obtained if $\frefl$ is larger:
if $\frefl=0.3$, then $\fhot/\fcooling\approx 0.35/0.33\approx 1.1$,
raising $\tau_{\rm H\,II}$ to $\sim3\times10^5 (0.5\cm^{-3}/n_e)\yr$,
but this is still short compared to O star lifetimes.
Longer values of $\tau_{\rm H\,II}$ can be obtained if $n_e$ is lowered,
but if $n_e$ is much smaller than $0.5\cm^{-3}$,
grain photoelectric heating and cosmic ray heating prevent the
gas from cooling to low enough temperature to be able to account
for the observed low free-free/H$\alpha$ ratio, if, as assumed, 
the cooling takes
place at $\sim$ constant density.

O star lifetimes are not the only time scale on which photoionization
can vary.
The ionizing radiation for the WIM may be provided in large part
by {\it runaway} O and B stars, with space velocities
$\gtsim 100\kms$.  Modulation of the photoionization rate
at a given point may result from motion of the ionizing sources
through a medium of varying opacity: a star moving at
$100\kms$ travels $10\pc$ in $10^5\yr$.

Although O stars have been generally 
favored as the source of
ionization for the WIM, 
it should be kept in mind that the actual source
for WIM ionization has not been securely established.
According to the present analysis, the ionization phase,
if it heats the plasma to $\sim10^4\K$,
needs to be of relatively short duration in order to be able to
explain the low observed ratio of free-free/H$\alpha$.
Perhaps other transient
processes (such as shock waves or soft X-rays due to supernova
blast waves, infalling
gas, etc.) contribute to heating and ionization of the WIM.

A prediction of the current model is that the H$\alpha$ emission
should have a component with the same radial velocity profile as the
21 cm emission, resulting from cosmic ray ionization of the
\hi. However, because cosmic ray ionization is thought to account for
$\leq2\%$ of the H$\alpha$, it may not be easy to recognize this
component. The reflected component of H$\alpha$, \nii, etc, should
also correlate with $N(\rm HI)$, but the apparent space velocities
will differ from 21 cm radial velocities due to motion of
the emitting \hii~relative to the reflecting dust grains.

\section{\label{sec:summary}
         Summary}

The principal points of this paper are as follows:

\begin{enumerate}
\item We simulated the cooling and recombination of initially
  photoionized gas in the WIM following removal of $h\nu >13.6$~eV
  radiation. The ratio of free-free emission to H$\alpha$, and
  emission line ratios such as \nii$\lambda$6583/H$\alpha$, were
  calculated, and various factors which influence the cooling were
  explored. The result strongly depends on the abundances of elements,
  gas density, cosmic ray ionization rate, and the abundance of very
  small grains and PAHs, but depends only weakly on the initial
  temperature and ionization fractions, as shown in
  Figure~\ref{zetanh}, \ref{major} and \ref{ic}.

\item Based on these calculations, we propose a three component model
  --- emission from hot ionized gas, cooling gas and neutral H\,I,
  plus reflected light -- to explain multiple observations
  in the WIM. With plausible weighting factors for these components,
  the model {\it simultaneously} yields the low free-free to H$\alpha$
  ratio, which indicates a low temperature ($\sim$3000\,K, DDF09), and
  metal line ratios, such as \nii$\lambda$6583/H$\alpha$ and
  \nii$\lambda$5755/\nii$\lambda$6583, which indicate a high
  temperature ($\sim10^4$\,K,
  \citet{Madsen+Reynolds+Haffner_2006}). The reflected component is
  crucial --- there must be a fraction ($\sim$20\%) of the observed
  H$\alpha$ coming from the reflected light, consistent with the
  estimate by \citet{Wood+Reynolds_1999}, to explain the observation.

\item For our model to successfully reproduce the low free-free/H$\alpha$ ratio,
some restrictions of the physical conditions in the WIM are required:
\begin{enumerate}
\item The ratio of cosmic ray primary ionization
rate to gas density $\zeta_{\rm CR}/n_{\rm H}$ should not exceed 
$\sim5\times10^{-16}$cm$^3$s$^{-1}$
(Figure~\ref{zetanh}), consistent with current observational estimates of $\zeta_{\rm CR}$ if $n_{\rm H}\approx0.5$cm$^{-3}$.
\item The gas phase element depletion parameter $F_\star\leq0.25$
  (Figure~\ref{major}) so that there are sufficient gas phase coolants
  to cool the gas in the presence of heating by cosmic rays and
  photoelectrons from grains, consistent with \citet{Jenkins_2009}.
\item The abundance of ultrasmall grains (including PAHs) should be suppressed by a factor $\sim$3 relative to the abundances in the overall \hi ~(Figure~\ref{major}), consistent with DDF09.
\end{enumerate}

\item Sightline-to-sightline variations of emission ratios
  are expected in our
  model. First, the parameters which affect the cooling model, such as
  the gas density, elemental abundances, cosmic ray ionization rate
  and the depletion factor of small grains, may vary spatially within
  the WIM.
  Second, different sightlines will have different proportions of
  the three components in our model.
  Third, the cooling component is time-dependent, and individual
  sightlines may not present a full average over the entire cooling
  history.
  These effects will lead to the variation of physical
  quantities deduced from observations on different sightlines, or
  for different velocity components on an individual
  sightline.

\end{enumerate}

\section*{Acknowledgments}

We thank R. Benjamin for helpful discussions, and the
anonymous referee for comments that led to improvements in the
manuscript. This research was
supported in part by NSF grants AST-0406883 and AST-1008570.

\begin{deluxetable}{lccccccccccccc}
\tabletypesize{\scriptsize}
\tablewidth{0pt}
\tablecaption{Initial Ionization Fractions}
\tablehead{
& &
\multicolumn{3}{c}{$IIF=1$\tablenotemark{a}} &&
\multicolumn{3}{c}{$IIF=2$\tablenotemark{b}} &&
\multicolumn{4}{c}{$IIF=3$\tablenotemark{c}}
\\
\cline{3-5} \cline{7-9} \cline{11-14} 
$A$ & $n_A/\nH$\tablenotemark{d} & I & II  & III   && I & II  & III  && I & II  & III & IV
}
\startdata
H & 1   & 0.05  & 0.95  & \nodata & & 0.05 & 0.95 & \nodata & & 0 & 1 & \nodata & \nodata \\
He& 0.1 &0.68  & 0.32  & 0 & & 0.5 & 0.5 & 0 & & 0.04 & 0.96 & 0 & \nodata \\
C & $10^{-3.65}$&0.01  & 0.94  & 0.05  & & 0.02 & 0.83 & 0.15 & & 0 & 0.13 & 0.87 & 0 \\
N & $10^{-4.21}$&0.05  & 0.93  & 0.02  & & 0.1 & 0.8 & 0.1 & & 0 & 0.13 & 0.87 & 0 \\
O & $10^{-3.25}$&0.05  & 0.95  & 0.00  & & 0.05 & 0.95 & 0.05 & & 0 & 0.43 & 0.57 & 0 \\
Ne& $10^{-4.03}$&0.11  & 0.89  & 0.00  & & 0.3 & 0.69 & 0.01 & & 0 & 0.91 & 0.09 & 0 \\
Mg& $10^{-4.65}$&0.01  & 0.70  & 0.29  & & 0.02 & 0.63 & 0.35 & & 0 & 0.05 & 0.95 & 0 \\
Si& $10^{-4.61}$&0.00  & 0.87  & 0.12  & & 0.01 & 0.74 & 0.25 & & 0 & 0.05 & 0.83 & 0.12 \\
S & $10^{-4.58}$&0.00  & 0.78  & 0.22  & & 0.02 & 0.68 & 0.3 & & 0 & 0.03 & 0.95 & 0.02 \\
Ar& $10^{-5.56}$&0.01  & 0.76  & 0.23  & & 0.15 & 0.75 & 0.1 & & 0 & 0.11 & 0.89 & 0 \\
Fe& $10^{-5.41}$&0.00  & 0.34  & 0.64  & & 0.01 & 0.69 & 0.3 & & 0 & 0.01 & 0.24 & 0.75 \\
\enddata
\tablenotetext{a}{Standard Model in \citet{Sembach+Howk+Ryans+Keenan_2000} with $\chi_{\rm edge}=0.1$}
\tablenotetext{b}{Estimate based on
\citet{Haffner+Reynolds+Tufte_1999, Reynolds_2004,
Madsen+Reynolds+Haffner_2006, Haffner+Dettmar+Beckman+etal_2009};
between IIF1 and IIF3}
\tablenotetext{c}{Orion nebula values \citep{Baldwin+Ferland+Martin+etal_1991}}
\tablenotetext{d}{Elemental abundance for standard Model ($F_\star=0$)}
\label{table:iif}
\end{deluxetable}
\begin{deluxetable}{ccccccccccc}
\tabletypesize{\scriptsize}
\tablewidth{0pc}
\tablecaption{Models for Recombining Gas}
\tablehead{
\colhead{Name\tablenotemark{a}} &
\colhead{$n_{\rm H}$\tablenotemark{b}} &
\colhead{$\zeta_{\rm CR}$\tablenotemark{c}} &
\colhead{$F_\star$\tablenotemark{d}} &
\colhead{g\tablenotemark{e}} &
\colhead{IIF\tablenotemark{f}} &
\colhead{$T_i$\tablenotemark{g}} &
\colhead{$\Psi$\tablenotemark{h}} &
\colhead{$\Phi$\tablenotemark{i}} &
\colhead{$\displaystyle\frac{\rm [NII]\lambda 5755}{\rm [NII]\lambda 6583}$ \tablenotemark{j} } &
\colhead{$\displaystyle\frac{\rm [SII]\lambda 6716}{\rm H\alpha}$ \tablenotemark{k} }
\\
 & \colhead{($cm^{-3}$)} & \colhead{($10^{-16}s^{-1}$)} & & & & \colhead{($K$)} & & & &
}

\startdata
standard & 0.5 & 1.0 & 0 & 0.33 & 1 & 8000 & 0.081 & 0.060 & 0.0039 & 0.15\\
$n_{\rm H}=2.5$ & 2.5 & 1.0 & 0 & 0.33 & 1 & 8000 & 0.080 & 0.063 & 0.0039 & 0.16\\
$n_{\rm H}=2.5\ \zeta_{\rm CR}=0.2$ & 2.5 & 0.2 & 0 & 0.33 & 1 & 8000 & 0.080 & 0.063 & 0.0039 & 0.15\\
$n_{\rm H}=2.5\ \zeta_{\rm CR}=2.0$ & 2.5 & 2.0 & 0 & 0.33 & 1 & 8000 & 0.080 & 0.062 & 0.0039 & 0.15\\
$n_{\rm H}=2.5\ \zeta_{\rm CR}=3.0$ & 2.5 & 3.0 & 0 & 0.33 & 1 & 8000 & 0.079 & 0.062 & 0.0039 & 0.15\\
$n_{\rm H}=2.5\ \zeta_{\rm CR}=5.0$ & 2.5 & 5.0 & 0 & 0.33 & 1 & 8000 & 0.079 & 0.061 & 0.0039 & 0.15\\
$\zeta_{\rm CR}=0.2$ & 0.5 & 0.2 & 0 & 0.33 & 1 & 8000 & 0.081 & 0.062 & 0.0039 & 0.15\\
$\zeta_{\rm CR}=2.0$ & 0.5 & 2.0 & 0 & 0.33 & 1 & 8000 & 0.083 & 0.059 & 0.0039 & 0.15\\
$\zeta_{\rm CR}=3.0$ & 0.5 & 3.0 & 0 & 0.33 & 1 & 8000 & 0.088 & 0.060 & 0.0039 & 0.15\\
$\zeta_{\rm CR}=5.0$ & 0.5 & 5.0 & 0 & 0.33 & 1 & 8000 & 0.111 & 0.078 & 0.0032 & 0.42\\
$F_\star=0.25$ & 0.5 & 1.0 & 0.25 & 0.33 & 1 & 8000 & 0.099 & 0.085 & 0.0040 & 0.14 \\
Reduced C & 0.5 & 1.0 & 0+C\tablenotemark{l} & 0.33 & 1 & 8000 & 0.088 & 0.065 & 0.0039 & 0.16 \\
$g=1$ & 0.5 & 1.0 & 0 & 1.0 & 1 & 8000 & 0.109 & 0.083 & 0.0038 & 0.27\\
$g=0.1$ & 0.5 & 1.0 & 0 & 0.1 & 1 & 8000 & 0.075 & 0.056 & 0.0039 & 0.13\\
IIF2 & 0.5 & 1.0 & 0 & 0.33 & 2 & 8000 & 0.083 & 0.054 & 0.0038 & 0.14\\
IIF3 & 0.5 & 1.0 & 0 & 0.33 & 3 & 8000 & 0.083 & 0.028 & 0.0028 & 0.063\\
$T_i=10^4$~K & 0.5 & 1.0 & 0 & 0.33 & 1 & 10000 & 0.086 & 0.099 & 0.0069 & 0.23\\
\enddata

\tablenotetext{a}{Model name}
\tablenotetext{b}{H nucleon density}
\tablenotetext{c}{Cosmic ray primary ionization rate. }
\tablenotetext{d}{Depletion parameter from \citep{Jenkins_2009}}
\tablenotetext{e}{The
grain reduction factor. The grain assisted recombination rate and
photoelectric heating rate reduce to this fraction of their full
values.}
\tablenotetext{f}{Initial ionization fractions (see Table~\ref{table:iif}) }
\tablenotetext{g}{Initial gas temperature.}
\tablenotetext{h}{Integrated ratio of free-free at 41 GHz to H$\alpha$, as in Equation~\ref{eq:psi}, for $n_{\rm H}t=10^7$cm$^{-3}$yr.}
\tablenotetext{i}{Integrated ratio of [NII]~$\lambda$6583 to H$\alpha$, as in Equation~\ref{eq:phi}, for $n_{\rm H}t=10^7$cm$^{-3}$yr.}
\tablenotetext{j}{Integrated ratio of [NII]~$\lambda$5755 to [NII]~$\lambda$6583, for $n_{\rm H}t=10^7$cm$^{-3}$yr.}
\tablenotetext{k}{Integrated ratio of [SII]~$\lambda$6716 to H$\alpha$, for $n_{\rm H}t=10^7$cm$^{-3}$yr.}
\tablenotetext{l}{Element abundance identical to the standard model
($F_\star=0$) except for C, for which we take the $F_\star=0$ value
$\times(2/3)$, or $n_{\rm C}/n_{\rm H}=1.4\times10^{-3}$} \label{table:models}
\end{deluxetable}


\begin{thebibliography}{70}
\expandafter\ifx\csname natexlab\endcsname\relax\def\natexlab#1{#1}\fi

\bibitem[{{Abrahamsson} {et~al.}(2007){Abrahamsson}, {Krems}, \&
  {Dalgarno}}]{Abrahamsson+Krems+Dalgarno_2007}
{Abrahamsson}, E., {Krems}, R.~V., \& {Dalgarno}, A. 2007, \apj, 654, 1171

\bibitem[{{Arnaud} \& {Raymond}(1992)}]{Arnaud+Raymond_1992}
{Arnaud}, M., \& {Raymond}, J. 1992, \apj, 398, 394

\bibitem[{{Asplund} {et~al.}(2009){Asplund}, {Grevesse}, {Sauval}, \&
  {Scott}}]{Asplund+Grevesse+Sauval+Scott_2009}
{Asplund}, M., {Grevesse}, N., {Sauval}, A.~J., \& {Scott}, P. 2009, \araa, 47,
  481

\bibitem[{{Bakes} \& {Tielens}(1994)}]{Bakes+Tielens_1994}
{Bakes}, E.~L.~O., \& {Tielens}, A.~G.~G.~M. 1994, \apj, 427, 822

\bibitem[{{Baldwin} {et~al.}(1991){Baldwin}, {Ferland}, {Martin}, {Corbin},
  {Cota}, {Peterson}, \& {Slettebak}}]{Baldwin+Ferland+Martin+etal_1991}
{Baldwin}, J.~A., {Ferland}, G.~J., {Martin}, P.~G., {Corbin}, M.~R., {Cota},
  S.~A., {Peterson}, B.~M., \& {Slettebak}, A. 1991, \apj, 374, 580

\bibitem[{{Barinovs} {et~al.}(2005){Barinovs}, {van Hemert}, {Krems}, \&
  {Dalgarno}}]{Barinovs+vanHemert+Krems+Dalgarno_2005}
{Barinovs}, {\u G}., {van Hemert}, M.~C., {Krems}, R., \& {Dalgarno}, A. 2005,
  \apj, 620, 537

\bibitem[{{Bautista} {et~al.}(2009){Bautista}, {Quinet}, {Palmeri}, {Badnell},
  {Dunn}, \& {Arav}}]{Bautista+Quinet+Palmeri+etal_2009}
{Bautista}, M.~A., {Quinet}, P., {Palmeri}, P., {Badnell}, N.~R., {Dunn}, J.,
  \& {Arav}, N. 2009, \aap, 508, 1527

\bibitem[{{Black} \& {van Dishoeck}(1991)}]{Black+vanDishoeck_1991}
{Black}, J.~H., \& {van Dishoeck}, E.~F. 1991, \apjl, 369, L9

\bibitem[{{Collins} \& {Rand}(2001)}]{Collins+Rand_2001}
{Collins}, J.~A., \& {Rand}, R.~J. 2001, \apj, 551, 57

\bibitem[{{Dalgarno} \& {McCray}(1972)}]{Dalgarno+McCray_1972}
{Dalgarno}, A., \& {McCray}, R.~A. 1972, \araa, 10, 375

\bibitem[{{Davies} {et~al.}(2006){Davies}, {Dickinson}, {Banday}, {Jaffe},
  {G{\'o}rski}, \& {Davis}}]{Davies+Dickinson+Banday+etal_2006}
{Davies}, R.~D., {Dickinson}, C., {Banday}, A.~J., {Jaffe}, T.~R.,
  {G{\'o}rski}, K.~M., \& {Davis}, R.~J. 2006, \mnras, 370, 1125

\bibitem[{{Dickey} {et~al.}(1978){Dickey}, {Terzian}, \&
  {Salpeter}}]{Dickey+Salpeter+Terzian_1978}
{Dickey}, J.~M., {Terzian}, Y., \& {Salpeter}, E.~E. 1978, \apjs, 36, 77

\bibitem[{{Dobler} {et~al.}(2009){Dobler}, {Draine}, \&
  {Finkbeiner}}]{Dobler+Draine+Finkbeiner_2009}
{Dobler}, G., {Draine}, B., \& {Finkbeiner}, D.~P. 2009, \apj, 699, 1374
(DDF09)

\bibitem[{{Dobler} \& {Finkbeiner}(2008)}]{Dobler+Finkbeiner_2008b}
{Dobler}, G., \& {Finkbeiner}, D.~P. 2008, \apj, 680, 1235

\bibitem[{{Draine}(1978)}]{Draine_1978}
{Draine}, B.~T. 1978, \apjs, 36, 595

\bibitem[{{Draine}(2011)}]{Draine_2011}
---. 2011, {Physics of the Interstellar and Intergalactic Medium} (Princeton,
  NJ: Princeton University Press)

\bibitem[{{Ferguson} {et~al.}(1996){Ferguson}, {Wyse}, {Gallagher}, \&
  {Hunter}}]{Ferguson+Wyse+Gallagher+Hunter_1996}
{Ferguson}, A.~M.~N., {Wyse}, R.~F.~G., {Gallagher}, III, J.~S., \& {Hunter},
  D.~A. 1996, \aj, 111, 2265

\bibitem[{{Ferri{\`e}re}(2001)}]{Ferriere_2001}
{Ferri{\`e}re}, K.~M. 2001, Reviews of Modern Physics, 73, 1031

\bibitem[{{Greenawalt} {et~al.}(1997){Greenawalt}, {Walterbos}, \&
  {Braun}}]{Greenawalt+Walterbos+Braun_1997}
{Greenawalt}, B., {Walterbos}, R.~A.~M., \& {Braun}, R. 1997, \apj, 483, 666

\bibitem[{{Griffin} {et~al.}(2001){Griffin}, {Mitnik}, \&
  {Badnell}}]{Griffin+Mitnik+Badnell_2001}
{Griffin}, D.~C., {Mitnik}, D.~M., \& {Badnell}, N.~R. 2001, \jphysb, 34, 4401

\bibitem[{{Haffner} {et~al.}(2009){Haffner}, {Dettmar}, {Beckman}, {Wood},
  {Slavin}, {Giammanco}, {Madsen}, {Zurita}, \&
  {Reynolds}}]{Haffner+Dettmar+Beckman+etal_2009}
{Haffner}, L.~M., {et~al.} 2009, Reviews of Modern Physics, 81, 969

\bibitem[{{Haffner} {et~al.}(1999){Haffner}, {Reynolds}, \&
  {Tufte}}]{Haffner+Reynolds+Tufte_1999}
{Haffner}, L.~M., {Reynolds}, R.~J., \& {Tufte}, S.~L. 1999, \apj, 523, 223

\bibitem[{{Hill} {et~al.}(2008){Hill}, {Benjamin}, {Kowal}, {Reynolds},
  {Haffner}, \& {Lazarian}}]{Hill+Benjamin+Kowal+etal_2008}
{Hill}, A.~S., {Benjamin}, R.~A., {Kowal}, G., {Reynolds}, R.~J., {Haffner},
  L.~M., \& {Lazarian}, A. 2008, \apj, 686, 363

\bibitem[{{Hoopes} {et~al.}(1996){Hoopes}, {Walterbos}, \&
  {Greenwalt}}]{Hoopes+Walterbos+Greenwalt_1996}
{Hoopes}, C.~G., {Walterbos}, R.~A.~M., \& {Greenwalt}, B.~E. 1996, \aj, 112,
  1429

\bibitem[{{Hoopes} {et~al.}(1999){Hoopes}, {Walterbos}, \&
  {Rand}}]{Hoopes+Walterbos+Rand_1999}
{Hoopes}, C.~G., {Walterbos}, R.~A.~M., \& {Rand}, R.~J. 1999, \apj, 522, 669

\bibitem[{{Hudson} \& {Bell}(2005)}]{Hudson+Bell_2005}
{Hudson}, C.~E., \& {Bell}, K.~L. 2005, \aap, 430, 725

\bibitem[{{Hummer}(1988)}]{Hummer_1988}
{Hummer}, D.~G. 1988, \apj, 327, 477

\bibitem[{{Indriolo} {et~al.}(2007){Indriolo}, {Geballe}, {Oka}, \&
  {McCall}}]{Indriolo+Geballe+Oka+McCall_2007}
{Indriolo}, N., {Geballe}, T.~R., {Oka}, T., \& {McCall}, B.~J. 2007, \apj,
  671, 1736

\bibitem[{{Jenkins}(2009)}]{Jenkins_2009}
{Jenkins}, E.~B. 2009, \apj, 700, 1299

\bibitem[{{Lepp}(1992)}]{Lepp_1992}
{Lepp}, S. 1992, in IAU Symp. 150: Astrochemistry of Cosmic Phenomena, 471--475

\bibitem[{{Madsen} {et~al.}(2006){Madsen}, {Reynolds}, \&
  {Haffner}}]{Madsen+Reynolds+Haffner_2006}
{Madsen}, G.~J., {Reynolds}, R.~J., \& {Haffner}, L.~M. 2006, \apj, 652, 401

\bibitem[{{Mathis} {et~al.}(1983){Mathis}, {Mezger}, \&
  {Panagia}}]{Mathis+Mezger+Panagia_1983}
{Mathis}, J.~S., {Mezger}, P.~G., \& {Panagia}, N. 1983, \aap, 128, 212

\bibitem[{{McCall} {et~al.}(2003){McCall}, {Huneycutt}, {Saykally}, {Geballe},
  {Djuric}, {Dunn}, {Semaniak}, {Novotny}, {Al-Khalili}, {Ehlerding},
  {Hellberg}, {Kalhori}, {Neau}, {Thomas}, {{\" O}sterdahl}, \&
  {Larsson}}]{McCall+Huneycutt+Saykally+etal_2003}
{McCall}, B.~J., {et~al.} 2003, \nature, 422, 500

\bibitem[{{McKee}(1990)}]{McKee_1990}
{McKee}, C.~F. 1990, in Astronomical Society of the Pacific Conference Series,
  Vol.~12, The Evolution of the Interstellar Medium, ed. {L.~Blitz}, 3--29

\bibitem[{{Miller} \& {Veilleux}(2003)}]{Miller+Veilleux_2003b}
{Miller}, S.~T., \& {Veilleux}, S. 2003, \apj, 592, 79

\bibitem[{{Nussbaumer} \& {Storey}(1986)}]{Nussbaumer+Storey_1986}
{Nussbaumer}, H., \& {Storey}, P.~J. 1986, \aaps, 64, 545

\bibitem[{{Otte} {et~al.}(2002){Otte}, {Gallagher}, \&
  {Reynolds}}]{Otte+Gallagher+Reynolds_2002}
{Otte}, B., {Gallagher}, III, J.~S., \& {Reynolds}, R.~J. 2002, \apj, 572, 823

\bibitem[{{Pequignot}(1996)}]{Pequignot_1996}
{Pequignot}, D. 1996, \aap, 313, 1026

\bibitem[{{Pequignot} {et~al.}(1991){Pequignot}, {Petitjean}, \&
  {Boisson}}]{Pequignot+Petitjean+Boisson_1991}
{Pequignot}, D., {Petitjean}, P., \& {Boisson}, C. 1991, \aap, 251, 680

\bibitem[{{Radhakrishnan} {et~al.}(1972){Radhakrishnan}, {Murray}, {Lockhart},
  \& {Whittle}}]{Radhakrishnan+Murray+Lockhart+Whittle_1972}
{Radhakrishnan}, V., {Murray}, J.~D., {Lockhart}, P., \& {Whittle}, R.~P.~J.
  1972, \apjs, 24, 15

\bibitem[{{Ramsbottom} {et~al.}(2007){Ramsbottom}, {Hudson}, {Norrington}, \&
  {Scott}}]{Ramsbottom+Hudson+Norrington+Scott_2007}
{Ramsbottom}, C.~A., {Hudson}, C.~E., {Norrington}, P.~H., \& {Scott}, M.~P.
  2007, \aap, 475, 765

\bibitem[{{Rand}(1996)}]{Rand_1996}
{Rand}, R.~J. 1996, \apj, 462, 712

\bibitem[{{Rand}(1997)}]{Rand_1997}
---. 1997, \apj, 474, 129

\bibitem[{{Rand}(2000)}]{Rand_2000}
---. 2000, \apjl, 537, L13

\bibitem[{{Rand} {et~al.}(1990){Rand}, {Kulkarni}, \&
  {Hester}}]{Rand+Kulkarni+Hester_1990}
{Rand}, R.~J., {Kulkarni}, S.~R., \& {Hester}, J.~J. 1990, \apjl, 352, L1

\bibitem[{{Reynolds}(1985)}]{Reynolds_1985}
{Reynolds}, R.~J. 1985, \apjl, 298, L27

\bibitem[{{Reynolds}(1991)}]{Reynolds_1991}
{Reynolds}, R.~J. 1991, in IAU Symposium, Vol. 144, The Interstellar Disk-Halo
  Connection in Galaxies, ed. {H.~Bloemen}, 67--76

\bibitem[{{Reynolds}(1993)}]{Reynolds_1993}
{Reynolds}, R.~J. 1993, in American Institute of Physics Conference Series,
  Vol. 278, Back to the Galaxy, ed. {S.~S.~Holt \& F.~Verter}, 156--165

\bibitem[{{Reynolds}(2004)}]{Reynolds_2004}
---. 2004, Advances in Space Research, 34, 27

\bibitem[{{Reynolds} {et~al.}(2001){Reynolds}, {Sterling}, {Haffner}, \&
  {Tufte}}]{Reynolds+Sterling+Haffner+Tufte_2001}
{Reynolds}, R.~J., {Sterling}, N.~C., {Haffner}, L.~M., \& {Tufte}, S.~L. 2001,
  \apjl, 548, L221

\bibitem[{{Rossa} \& {Dettmar}(2000)}]{Rossa+Dettmar_2000}
{Rossa}, J., \& {Dettmar}, R. 2000, \aap, 359, 433

\bibitem[{{Savage} \& {Sembach}(1996)}]{Savage+Sembach_1996}
{Savage}, B.~D., \& {Sembach}, K.~R. 1996, \araa, 34, 279

\bibitem[{{Sembach} {et~al.}(2000){Sembach}, {Howk}, {Ryans}, \&
  {Keenan}}]{Sembach+Howk+Ryans+Keenan_2000}
{Sembach}, K.~R., {Howk}, J.~C., {Ryans}, R.~S.~I., \& {Keenan}, F.~P. 2000,
  \apj, 528, 310

\bibitem[{{Shull} \& {van Steenberg}(1982)}]{Shull+vanSteenberg_1982}
{Shull}, J.~M., \& {van Steenberg}, M. 1982, \apjs, 48, 95

\bibitem[{{Sofia} \& {Parvathi}(2010)}]{Sofia+Parvathi_2010}
{Sofia}, U.~J., \& {Parvathi}, V.~S. 2010, in Cosmic Dust -- Near and Far, ed.
  T.~{Henning}, E.~{Gr\"un}, \& J.~{Steinacker}, 236--242

\bibitem[{{Spitzer}(1985)}]{Spitzer_1985}
{Spitzer}, Jr., L. 1985, \apjl, 290, L21

\bibitem[{{Stancil} {et~al.}(1999){Stancil}, {Schultz}, {Kimura}, {Gu},
  {Hirsch}, \& {Buenker}}]{Stancil+Schultz+Kimura+etal_1999}
{Stancil}, P.~C., {Schultz}, D.~R., {Kimura}, M., {Gu}, J.-P., {Hirsch}, G., \&
  {Buenker}, R.~J. 1999, \aaps, 140, 225

\bibitem[{{Tayal}(2007)}]{Tayal_2007}
{Tayal}, S.~S. 2007, \apjs, 171, 331

\bibitem[{{Tayal}(2008)}]{Tayal_2008b}
---. 2008, \aap, 486, 629

\bibitem[{{Tayal} \& {Gupta}(1999)}]{Tayal+Gupta_1999}
{Tayal}, S.~S., \& {Gupta}, G.~P. 1999, \apj, 526, 544

\bibitem[{{Tayal} \& {Zatsarinny}(2010)}]{Tayal+Zatsarinny_2010}
{Tayal}, S.~S., \& {Zatsarinny}, O. 2010, \apjs, 188, 32

\bibitem[{{Verner}(1999)}]{Verner_1999}
{Verner}, D.~A. 1999, Subroutine rrfit, version 4,
  http:///www.pa.uky.edu/$\sim$verner/fortran.html

\bibitem[{{Verner} \& {Ferland}(1996)}]{Verner+Ferland_1996}
{Verner}, D.~A., \& {Ferland}, G.~J. 1996, \apjs, 103, 467

\bibitem[{{Verner} {et~al.}(1996){Verner}, {Ferland}, {Korista}, \&
  {Yakovlev}}]{Verner+Ferland+Korista+Yakovlev_1996}
{Verner}, D.~A., {Ferland}, G.~J., {Korista}, K.~T., \& {Yakovlev}, D.~G. 1996,
  \apj, 465, 487

\bibitem[{{Verner} \& {Yakovlev}(1995)}]{Verner+Yakovlev_1995}
{Verner}, D.~A., \& {Yakovlev}, D.~G. 1995, \aaps, 109, 125

\bibitem[{{Wang} {et~al.}(2002){Wang}, {Seo}, {Anraku}, {Fujikawa}, {Imori},
  {Maeno}, {Matsui}, {Matsunaga}, {Motoki}, {Orito}, {Saeki}, {Sanuki}, {Ueda},
  {Yoshimura}, {Makida}, {Suzuki}, {Tanaka}, {Yamamoto}, {Yoshida}, {Mitsui},
  {Matsumoto}, {Nozaki}, {Sasaki}, {Mitchell}, {Moiseev}, {Ormes},
  {Streitmatter}, {Nishimura}, {Yajima}, \&
  {Yamagami}}]{Wang+Seo+Anraku+etal_2002}
{Wang}, J.~Z., {et~al.} 2002, \apj, 564, 244

\bibitem[{{Webber} \& {Yushak}(1983)}]{Webber+Yushak_1983}
{Webber}, W.~R., \& {Yushak}, S.~M. 1983, \apj, 275, 391

\bibitem[{{Weingartner} \&
  {Draine}(2001{\natexlab{a}})}]{Weingartner+Draine_2001d}
{Weingartner}, J.~C., \& {Draine}, B.~T. 2001{\natexlab{a}}, \apj, 563, 842

\bibitem[{{Weingartner} \&
  {Draine}(2001{\natexlab{b}})}]{Weingartner+Draine_2001c}
---. 2001{\natexlab{b}}, \apjs, 134, 263

\bibitem[{{Wood} \& {Reynolds}(1999)}]{Wood+Reynolds_1999}
{Wood}, K., \& {Reynolds}, R.~J. 1999, \apj, 525, 799

\end{thebibliography}
\end{document}